\documentclass[11pt,a4paper,notoc]{article}
\usepackage{jheppub}
\usepackage{amsmath}         
\usepackage{amssymb}          
\usepackage{bbold}
\usepackage{xcolor}
\usepackage{ulem}
\usepackage{afterpage}
\usepackage{youngtab}
\usepackage{multirow}
\usepackage[capitalise]{cleveref}
\usepackage{slashed}
\usepackage{booktabs}
\usepackage{subcaption}
\usepackage{comment}
\usepackage{rotating}
\usepackage{url}
\usepackage{physics}
\usepackage{slashbox}

\crefname{section}{sec.}{secs.}
\crefname{table}{Tab.}{Tabs.}
\crefname{figure}{Fig.}{Figs.}
\crefname{equation}{Eq.}{Eqs.}
\crefname{appendix}{Appendix}{Appendix}

\newcommand{\hc}{\text{h.c.}}
\newcommand{\SO}{\text{SO}}
\newcommand{\SU}{\text{SU}}
\newcommand{\U}{\text{U}}
\newcommand{\Sp}{\text{Sp}}

\newcommand{\br}{\mathrm{Br}}

\newcommand{\lr}{\overset{\leftrightarrow}{\partial}}

\def\beq{\begin{equation}}
\def\eeq{\end{equation}}

\def\gsim{\raise0.3ex\hbox{$\;>$\kern-0.75em\raise-1.1ex\hbox{$\sim\;$}}}
\def\lsim{\raise0.3ex\hbox{$\;<$\kern-0.75em\raise-1.1ex\hbox{$\sim\;$}}}

\usepackage[utf8]{inputenc}

\title{Coloured spin-1 states in composite Higgs models}

\author[a,b]{Giacomo~Cacciapaglia,}
\affiliation[a]{Universite Claude Bernard Lyon 1, CNRS/IN2P3, IP2I UMR 5822,  4 rue Enrico Fermi, F-69100 Villeurbanne, France}
\affiliation[b]{Quantum Theory Center (QTC) \& D-IAS, Southern Denmark Univ., Campusvej 55, 5230 Odense M, Denmark}
\emailAdd{g.cacciapaglia@ip2i.in2p3.fr}

\author[a,c]{Aldo~Deandrea,}
\affiliation[c]{Department of Physics, University of Johannesburg, PO Box 524, Auckland Park 2006, South Africa}
\emailAdd{deandrea@ip2i.in2p3.fr}

\author[d]{Manuel~Kunkel,}
\affiliation[d]{Institut f\"{u}r Theoretische Physik und Astrophysik, Uni W\"{u}rzburg, Emil-Hilb-Weg 22, D-97074 W\"{u}rzburg, Germany}
\emailAdd{manuel.kunkel@uni-wuerzburg.de}

\author[d]{Werner~Porod}
\emailAdd{porod@physik.uni-wuerzburg.de}

\begin{document}

\abstract{Strong dynamics for composite Higgs models predict spin-1 resonances
which are expected to be in the same mass range as the usually considered
top-partners.
We study here QCD-coloured vector and axial-vector states stemming from
composite Higgs dynamics in several relevant
models based on an underlying gauge-fermion description. These states
can come as  triplet, sextet and octet representation. All models
considered have a colour octet vector state in common which can be singly
produced at hadron colliders as it mixes with the gluon.
 We
explore the rich and testable phenomenology of these coloured spin-1
states at the LHC and future colliders.}

\maketitle

\section{Introduction}

The exploration of composite Higgs models has garnered significant attention in the realm of theoretical particle physics, as these models offer a possible explanation for the nature of the Higgs boson discovered at CERN and a dynamical origin for the breaking of the electroweak symmetry in the Standard Model (SM) \cite{Englert:1964et,Higgs:1964pj,Guralnik:1964eu}. By positing the Higgs boson as a composite state that originates from a new strongly interacting sector, composite Higgs models provide a potential solution to the problem of hierarchy between the electroweak scale and the Planck scale: like in quantum chromodynamics (QCD), the breaking scale is dynamically generated via confinement and condensation of a new interaction. This idea is as old as the SM itself \cite{Weinberg:1975gm,Dimopoulos:1979es}, starting from the first Higgsless (Technicolor) theories \cite{Dimopoulos:1979za} and their effective Lagrangian counterparts \cite{Casalbuoni:1985kq}, to models where the Higgs emerges as a meson \cite{Kaplan:1983fs,Kaplan:1983sm}. Composite model building has resumed in the early 2000's thanks to the idea of holography \cite{Contino:2003ve,Agashe:2004rs,Hosotani:2005nz}, freely adapted from supersymmetric string theory inspired duality conjectures \cite{Maldacena:1997re}.

The varied and rich phenomenology of composite Higgs models has been extensively studied, both from the point of view of holography-inspired effective models \cite{Contino:2010rs,Panico:2015jxa} and from models based on underlying gauge-fermion theories \cite{Cacciapaglia:2020kgq}, the latter close in spirit to QCD. While we do not attempt to summarise the main features, which have been described several times, we want to recall the essential ingredients of such theories, which are directly related to the electroweak symmetry breaking. The Higgs typically emerges as a pseudo-Nambu Goldstone boson (pNGB) \cite{Contino:2003ve} from the spontaneous breaking of the global symmetry in the strong sector. Its potential and mass are generated by explicit breaking terms: the gauging of the electroweak symmetry, the couplings of the top quark \cite{Agashe:2004rs} and (eventually) a mass term for the underlying fermions \cite{Galloway:2010bp,Cacciapaglia:2014uja}. In this framework, composite spin-1  states matching the electroweak gauge bosons and top partners have been widely considered. From the Higgs sector point of view, they are the minimal components required from the strong sector.

Nevertheless, the strong dynamics of composite Higgs models is much richer than this. Whether it consists of an unspecified conformal field theory in the holographic approach, or of a well-defined gauge-fermion theory, a more extended spectrum is a generic prediction. In particular, the fact that top partners \cite{Kaplan:1991dc} need to be charged under QCD interactions implies that other coloured resonances beyond the top partners must exist. This implies the presence of coloured spin-0 and spin-1 mesons, as well as fermions carrying unusual colour charges. In this work we will focus on coloured spin-1 resonances, which are expected to exist in all types of composite Higgs models. In holographic models, they emerge as Kaluza-Klein resonances of the gluon field \cite{Guchait:2007ux}. As we will show, however, a richer set of coloured spin-1 states is to be expected.

For definiteness, we will focus on theories based on an underlying gauge-fermion description, where the properties and quantum numbers of the resonances can be classified. A systematic list of models describing the minimal resonances needed by the Higgs sector has been presented by Ferretti and Karateev \cite{Ferretti:2013kya}. Consistent models with a single species of fermions can only be based on $\SU(3)$ \cite{Vecchi:2015fma} -- like in QCD -- or $G_2$ \cite{Ferretti:2013kya} with fermions in the fundamental. However, models with two separate species in different irreducible representations (irreps) of the gauge group offer the intriguing possibility of sequestering QCD interactions from the sector responsible for the electroweak symmetry breaking \cite{Barnard:2013zea,Ferretti:2013kya}. Theoretical and phenomenological considerations lead to the definition of 12 minimal models, whose characteristics are fully specified \cite{Ferretti:2016upr,Belyaev:2016ftv} in terms of the confining gauge group and the irreps and multiplicities of the two species of fermions. Upon confinement, both fermion species condense, as confirmed by Lattice results for $\SU(4)$ and $\Sp(4)$ gauge symmetries \cite{Ayyar:2017uqh,Bennett:2023wjw}, hence generating two sets of pNGBs \cite{Ferretti:2016upr} (plus one coming from a global anomaly-free U(1) \cite{Belyaev:2016ftv}). The symmetry breaking patterns are uniquely determined by the type of irrep the two species belong to \cite{Cacciapaglia:2014uja}, leading to the classification in Table~\ref{tab:classification}. The top partners emerge as so called ``chimera'' baryons formed of the two species, where two different patterns can be realised: $\psi \psi \chi$ and $\psi \chi \chi$, where $\psi$ only carry electroweak charges while $\chi$ carry QCD colour and hypercharge. In the former case, the $\chi$'s QCD triplet carries hypercharge $2/3$, in the latter case $-1/3$.

\begin{table}[htb]
\centering
\begin{tabular}{|l||c|c|c|}
\hline
\backslashbox{QCD}{EW} & SU(4)/Sp(4) & SU(5)/SO(5) & SU(4)$^2$/SU(4) \\ \hline\hline
\multirow{2}{*}{SU(6)/Sp(6)} & &  & \\
& & M5 ($\psi\chi\chi$) & \\ \hline
\multirow{2}{*}{SU(6)/SO(6)} & M8-9 ($\psi\psi\chi$) & M3-4 ($\psi\psi\chi$)  & M10-11 ($\psi\psi\chi$) \\ 
&  & M1-2 ($\psi\chi\chi$) & \\\hline
\multirow{2}{*}{SU(3)$^2$/SU(3)} & &  & M12 ($\psi\psi\chi$)\\
 & & M6-7 ($\psi\chi\chi$) & \\ \hline
\end{tabular}
\caption{\label{tab:classification} Classification of the 12 models based on the global symmetry breaking patterns in the electroweak and QCD sectors. In parenthesis, we indicate the template for the chimera baryons, representing the top partners.}
\end{table}

The phenomenology of the resonances from these 12 models have been studied in the literature, covering some of the resonance types. So far, studies have focused on the  pNGBs  charged under electroweak quantum numbers \cite{Ferretti:2016upr,Agugliaro:2018vsu,Cacciapaglia:2022bax}, the singlets stemming from the global $\U(1)$'s \cite{Ferretti:2016upr,Belyaev:2016ftv,Cacciapaglia:2019bqz,BuarqueFranzosi:2021kky}, QCD coloured pNGBs \cite{Cacciapaglia:2015eqa,Belyaev:2016ftv,Cacciapaglia:2020vyf}, top partners with non-standard decays \cite{Bizot:2018tds,Xie:2019gya,Cacciapaglia:2019zmj} or colour assignment \cite{Cacciapaglia:2021uqh}, and spin-1 resonances carrying electroweak charges \cite{BuarqueFranzosi:2016ooy}. We also note that the spectra and couplings of such resonances can be computed on the Lattice, and some results are available for models based on $\Sp(4)$ \cite{Bennett:2017kga,Bennett:2019cxd,Bennett:2019jzz,Bennett:2020hqd,Bennett:2020qtj,Bennett:2022yfa,Bennett:2023gbe,Bennett:2023mhh}, like models M5 and M8, and based on $\SU(4)$ \cite{Ayyar:2017qdf,Ayyar:2018glg,Ayyar:2018ppa,Ayyar:2018zuk,Ayyar:2019exp,Golterman:2020pyx,Hasenfratz:2023sqa}, like models M6 and M11. Computations based on holography are also available \cite{Erdmenger:2020lvq,Erdmenger:2020flu,Elander:2020nyd,Elander:2021bmt,Erdmenger:2023hkl,Erdmenger:2024dxf}.

In this work, we will focus on the phenomenology of spin-1 resonances that carry QCD charges and emerge as bound states of the $\chi$ species. Their properties emerge from three types of cosets, $\SU(6)/\SO(6)$, $\SU(6)/\Sp(6)$ and $\SU(3)\times \SU(3)/\SU(3)$, and the hypercharge assignment for the colour triplet $\chi$, which stems from the types of chimera baryons. 
The spectrum contains both a set of vectors $\mathcal V^\mu$ and of axial-vectors $\mathcal A^\mu$, which decay respectively into two or three pNGBs. The latter property originates from the symmetric nature of the cosets. 
Mixing of the ubiquitous octet with the QCD gluons will also generate direct couplings to quarks, while the colour triplets and sextets may or may not couple to a pair of quarks depending on their baryon number. To properly characterise the phenomenology of these states, we will employ the hidden symmetry approach \cite{Bando:1987br} to write an effective Lagrangian, and use the results to study their collider phenomenology.

The paper is structured as follows: in Section~\ref{sec:hidden} we briefly review the hidden symmetry approach and present results in the allowed cases. In Section~\ref{sec:pheno} we analyse the phenomenology at the LHC and future high energy proton colliders. Finally, we offer our conclusions in Section~\ref{sec:conclu}.

\section{Hidden gauge symmetry approach} \label{sec:hidden}

The hidden local symmetry method is based on the idea that the nonlinear $\sigma$ model on the manifold $G/H$ is gauge equivalent to the $\sigma$ model based on $G \times H_\mathrm{local}$. The
gauge bosons corresponding to the local symmetry can be identified with composite spin-1 mesons. The general procedure for building an effective Lagrangian including these new spin-1 resonances \cite{Bando:1984ej,Casalbuoni:1985kq,Casalbuoni:1988xm} consists, therefore, in a generalised group structure that splits the unitary matrix $U(x)$ describing the Goldstone bosons into factors  
transforming under an extended symmetry $G'$.

The generators of the group $G$ can be indicated with $T^A$ where $A=1, \ldots, d_G$ and $d_G$ is the dimension of the group $G$. These generators can be separated into two classes, $T^A = \{ S^a, X^I\}$: the unbroken generators $S^a$ with $a=1, \ldots, d_H$ belonging to the unbroken subgroup $H \subset G$, and the broken generators $X^{I}$ with $I=1, \ldots, d_G - d_H$ belonging to the coset $G/H$. The elements of $G$ are of the form $g=e^{i \alpha^A T^A}$ and those of $H$ of the form $h=e^{i \beta^a S^a}$. The elements of
$G$ can be parameterised by $g = U h$ with $U$ in the coset $G/H$
\begin{equation}
   U   = e^{i \pi^I X^I}\; .
\end{equation}
For cosets of the type $\SU(N)/\SO(N)$, $\SU(2N)/\Sp(2N)$ and $\SU(N)_L \times \SU(N)_R/\SU(N)_V$,
the two classes of generators are determined by the following constraints:
\begin{equation}
S^a \Sigma_0 + \Sigma_0 S^{a T} = 0 \;, \;\;\; X^{I} \Sigma_0 - \Sigma_0 X^{IT} = 0,
\end{equation}
see \cite{Peskin:1980gc,Preskill:1980mz,BuarqueFranzosi:2023xux} for details. The Lagrangian in the condensate (``chiral") phase is built using the standard chiral Lagrangian elements:
\begin{equation}
\Omega_\mu = iU^\dagger D_\mu U \;, \;\;\; D_\mu  U=(\partial_\mu -i\, j_\mu )\, U\;, 
\end{equation}
with $\Omega_\mu$ the Maurer-Cartan form and $j_\mu$ the current $j_\mu = v^a_\mu S^a +a^{I}_\mu X^{I}$. 
The form $\Omega_\mu$ can be further decomposed into projections along the unbroken and broken parts:
\begin{align}
	e_{\mu} &= 2 \Tr(S^a \Omega_{\mu})\ S^a\,, \label{eq:e_definition}\\
	d_{\mu} &= 2 \Tr(X^{I} \Omega_{\mu})\ X^{I}\,, \label{eq:d_definition}
\end{align}
which will be explicitly used in writing the Lagrangian.
The notation for the current $j_\mu$ indicates that vector resonances $v^a_\mu$ are associated to the unbroken generators of $H$, while axial-vectors $a^{I}_\mu$ to the broken ones. This is a formal definition, while a direct correspondence to vector and axial currents of fermions is only recovered in QCD-like cases based on $\SU(N)_L \times \SU(N)_R$ group symmetries.

For concreteness, in the rest of the section we will provide some details on the effective construction for one of the cosets, based on $\SU(6)/\SO(6)$. We will show how to extend the results to the other two cosets (c.f. \cref{tab:classification}) at the end of the section.

\subsection{Setup for $\SU(6)/\SO(6)$}

Following the hidden symmetry prescriptions, we consider a model based on the symmetry $G' = \SU(6)_0\times \SU(6)_1$, where $\SU(6)_0$ is partly gauged by the SM gauge bosons (gluons and hypercharge) and $\SU(6)_1$ is fully gauged by the heavy resonances.
The enlarged symmetry is broken to $\SO(6)_0\times\SO(6)_1$ by two sets of pNGBs, $\pi_0$ and $\pi_1$, so that a linear combination of them gives mass to the axial resonances.
Furthermore, $\SO(6)_0 \times \SO(6)_1$ is broken to the diagonal subgroup $\SO(6)$ by a second set of pNGBS, $k$, which gives mass to the vector resonances.

We parameterise the two sets of pNGBs as
\begin{equation}
	U_0 = \exp[ \frac{i\sqrt 2}{f_0} \, \pi_0^I X^I ] \quad \text{and{ \quad}} 	U_1 = \exp[ \frac{i\sqrt 2}{f_1} \, \pi_1^I X^I ]\,,
\end{equation}
transforming under $\SU(6)_i$ as
\begin{equation}
	U_i \to g_i U_i h(g_i, \pi_i)^\dagger .
\end{equation}
We also define a Maurer-Cartan form for each sector:
\begin{equation}
	\Omega_{i,\mu} = iU_i^\dagger D_\mu U_i\,,
\end{equation}
with covariant derivatives
\begin{align}
	D_\mu U_0 &= ( \partial_\mu -i\hat{g}_s \mathbf  G_\mu - i\hat{g}' \mathbf  B_\mu ) \, U_0\,, \\
	D_\mu U_1 &= ( \partial_\mu -i\tilde g \boldsymbol{\mathcal V}_\mu - i\tilde g \boldsymbol{\mathcal A}_\mu )\, U_1\,,
\end{align}
where the gauge fields act via the commutator, $[\mathbf G_\mu,U_0]$ etc, and
\begin{align}
	\mathbf  B_\mu = B_\mu\, T_X, \quad \mathbf  G_\mu = G^a_\mu\, T_G^a, \quad \boldsymbol{\mathcal V}_\mu = \mathcal V_\mu^a\, S^a, \quad \boldsymbol{\mathcal A}_\mu = \mathcal A^I_\mu \, X^I\,,
\end{align}
where $T_X$ and $T_G^a$ are the generators of $\SO(6)_0$ corresponding to hypercharge and QCD colour, respectively.
The colour multiplets are embedded in the $\SO(6)$ matrices as
\begin{gather}
    \boldsymbol \pi= \frac{1}{\sqrt 2}\mqty( \boldsymbol \pi_8 & \boldsymbol \pi_6 \\ \boldsymbol \pi_6^c & \boldsymbol \pi_8^T ), \qquad
	\boldsymbol {\mathcal A}^\mu = \frac{1}{\sqrt 2} \mqty( \boldsymbol {\mathcal A}_8^\mu & \boldsymbol {\mathcal A}_6^\mu \\ \boldsymbol {\mathcal A}_6^{c,\mu} & \boldsymbol {\mathcal A}_8^{\mu,T} ) ,\\
	\boldsymbol {\mathcal V}^\mu = \frac{1}{\sqrt 2} \mqty( \boldsymbol {\mathcal V}_8^\mu + \frac{1}{\sqrt 6} \boldsymbol {\mathcal V}_1^\mu & \boldsymbol {\mathcal V}_3^{c,\mu} \\ \boldsymbol {\mathcal V}_3^\mu  & -\boldsymbol {\mathcal V}_8^{\mu,T} - \frac{1}{\sqrt 6} \boldsymbol {\mathcal V}_1^\mu ) ,
\end{gather}
where $\boldsymbol \phi_8=\frac 12 \phi_8^a \lambda^a$ with the Gell-Mann matrices $\lambda^a$, $\boldsymbol \phi_{6}=\boldsymbol \phi_6^T$, and $\boldsymbol \phi_3=-\boldsymbol \phi_3^T$.
To employ the CCWZ construction \cite{Coleman:1969sm,Callan:1969sn}, we define the components of the Maurer-Cartan forms $d_{i,\mu}$ and $e_{i,\mu}$ parallel and orthogonal to $\SO(6)_i$ as in \cref{eq:e_definition,eq:d_definition}.
They transform under $\SU(6)_i$ as
\begin{align}
	d_{i,\mu} &\to h(g_i, \pi_i) \,d_{i,\mu}\, h^\dagger (g_i, \pi_i)\,, \\
	e_{i,\mu} &\to h(g_i, \pi_i) (e_{i,\mu} + i\partial_\mu) h^\dagger (g_i, \pi_i)\,. 
\end{align}
We refer the reader to \cref{app:ccwz} for the explicit calculation of the CCWZ symbols.

For the $\SO(6)_0\times\SO(6)_1 \to \SO(6)$ breaking we introduce a second set of pNGBs
\begin{equation}
	K = \exp[ \frac{i}{f_K} k^a\, S^a ]
\end{equation}
transforming as:
\begin{equation}
	K \to h(g_0, \pi_0) \,K\, h^\dagger (g_1, \pi_1)
\end{equation}
with covariant derivative:
\begin{equation}
	D_\mu K = \partial_\mu K - i e_{0,\mu} K + iK e_{1,\mu}\,.
\end{equation}

\subsection{The Lagrangian}
From the previous considerations, and in a similar way to what was obtained in the corresponding $\SU(4)$ case in \cite{BuarqueFranzosi:2016ooy}, the most general, leading-order Lagrangian reads:
\begin{align} \label{eq:Lagr}
	\mathcal L = &-\frac{1}{2} \Tr \mathbf  G_{\mu\nu} \mathbf  G^{\mu\nu} - \frac{1}{2} \Tr \mathbf  B_{\mu\nu} \mathbf  B^{\mu\nu} - \frac{1}{2} \Tr \boldsymbol{\mathcal F}_{\mu\nu} \boldsymbol{\mathcal F}^{\mu\nu} \nonumber \\
	&+ \frac{f_0^2}{2} \Tr d_{0,\mu} d_0^{\mu} + \frac{f_1^2}{2} \Tr d_{1,\mu} d_1^{\mu} \nonumber \\
	&+ \frac{f^2_K}{2} \Tr D^\mu K (D_\mu K)^\dagger + rf_1^2 \, \Tr d_{0,\mu} K d^{\mu}_1 K^\dagger \nonumber\\
	&+ \mathcal L_\mathrm{fermions}
\end{align}
where 
\begin{equation}
	\boldsymbol{\mathcal F}_\mu = \boldsymbol{\mathcal V}_\mu + \boldsymbol{\mathcal A}_\mu
\end{equation}
contains all the massive resonances.
We recall that for a generic gauge field $\mathbf  V_\mu$,
\begin{equation}
	\mathbf  V_{\mu\nu} = \partial_\mu \mathbf  V_\nu - \partial_\nu \mathbf  V_\mu - i g [\mathbf  V_\mu, \mathbf  V_\nu]\,,
\end{equation}
where $g$ is the appropriate coupling.
In the unitary gauge, where $K = \mathbb{1}$, the kinetic term for $K$ simplifies to
\begin{align}\label{eq:Kkin}
	\Tr D_\mu K (D^\mu K)^\dagger &= \Tr e_{0,\mu} e_0^{\mu} + \Tr e_{1,\mu} e_1^{\mu} - 2\Tr e_{0,\mu} e_1^{\mu} \,.
\end{align}
Expanding the above Lagrangian will allow us to compute the mass eigenstates (elementary vectors and resonances do mix) and their couplings.

\subsection{Vector boson masses and mixing} 

The masses and mixing of the vector resonances stem from the pNGB matrix $K$.
The three terms in \cref{eq:Kkin} read
\begin{align}
	\Tr(e_{0,\mu} e_0^{\mu}) &\supset \hat{g}_s^2 \Tr(\mathbf G_\mu \mathbf G^\mu  ) + \hat{g}^{\prime \, 2} \Tr(\mathbf B_\mu \mathbf B^\mu) = \frac{\hat{g}^{2}_s}{2} G^a_\mu G^{\mu,a} + \frac{\hat{g}^{\prime\,2}}{2} B_\mu B^\mu\,, \\
	\Tr(e_{1,\mu} e_1^{\mu}) &\supset \tilde g^2 \Tr(\boldsymbol {\mathcal V}_\mu \boldsymbol {\mathcal V}^\mu ) = \frac 12 \tilde g^2 \mathcal V_\mu^a \mathcal V^{a,\mu}\,,  \label{eq:kin_V} \\
	\Tr(e_{0,\mu} e_1^{\mu}) &\supset \Tr((\hat{g}_s \mathbf G_\mu + \hat{g}'\mathbf B_\mu) \tilde g \boldsymbol{\mathcal V}^\mu ))  = \frac 12 \hat{g}_s \tilde g\, G^a_\mu \mathcal V_8^{a,\mu} + \frac 12 \hat{g}' \tilde g \, B_\mu \mathcal V_1^\mu\,,
\end{align}
where, from the last line, we see that the colour octet and singlet components mix with gluons and the hypercharge gauge boson, respectively.

The Lagrangian contains a simple mass term for the colour-triplet state:  
\begin{equation}
	M_{\mathcal V_{3}}= \frac{\tilde g f_K}{\sqrt 2}.
\end{equation}
For the other states, a mixed mass term emerges. Starting with the colour octets:
\begin{align}
	\mathcal L &\supset  \frac{f_K^2}{4} \tilde g^2\, \mathcal V^a_{8,\mu} \mathcal V_8^{a,\mu} + \frac{f_K^2}{4} \hat{g}_s^2\, G^a_\mu G^{a,\mu} - \frac{f_K^2}{2} \hat{g}_s \tilde g \, G^a_\mu \mathcal V_8^{a,\mu} =\frac 12 V_{8,\mu}^{a,T} \ \mathcal M_8^2\ V^{a,\mu}_8\,, 
\end{align}
where 
\begin{equation}
	V^a_{8,\mu} = \mqty(G^a_\mu \\ \mathcal V^a_{8,\mu}), \qquad \mathcal M_8^2 = \frac{f_K^2}{2} \mqty(\hat{g}_s^2 & -\hat{g}_s\tilde g \\ -\hat{g}_s\tilde g & \tilde g^2)\,.
\end{equation}
Diagonalising the mass matrix, we find a massless eigenstate, which corresponds to the physical gluon octet, and a massive state, corresponding to the octet vector resonance. The latter have mass
\begin{equation}
    M_{\mathcal V_8} = \frac{f_K}{\sqrt 2} \sqrt{\tilde g^2 + \hat{g}_s^2}\,.
\end{equation}
With some abuse of notation, we can switch to the physical mass eigenstates by replacing
\begin{equation}
	\mqty(G^a_\mu \\ \mathcal V^a_{8,\mu}) \to \mqty(\cos\beta_8 & -\sin\beta_8 \\ \sin\beta_8 & \cos\beta_8) \mqty(G^a_\mu \\ \mathcal V^a_{8,\mu}), \qquad  \tan\beta_8 = \frac{\hat{g}_s}{\tilde g} \lesssim 1\,.
\end{equation}
Finally, the gauge coupling associated to the massless gluons reads
\begin{equation} \label{eq:gsphys}
    g_s = \hat{g}_s \cos \beta_8 = \tilde{g} \sin \beta_8 = \frac{\hat{g}_s \tilde{g}}{\sqrt{\hat{g}_s^2 + \tilde{g}^2}}\,,
\end{equation}
and this corresponds to the physical coupling of QCD interactions.

A similar mixing pattern emerges in the singlet, leading to 
\begin{equation}
	M_{\mathcal V_1} = \frac{f_K}{\sqrt 2} \sqrt{\tilde g^2 + \hat{g}^{\prime\,2}},\qquad \tan\beta_1 = \frac{\hat{g}'}{\tilde g}\,,
\end{equation}
with the caveat that the hypercharge will also mix with a spin-1 resonance stemming from the electroweak sector of the composite theory. Such a mixing has been studied for the $\SU(4)/\Sp(4)$ coset in \cite{BuarqueFranzosi:2016ooy}. Combining the two sectors will, therefore, lead to a more complicated mixing pattern. We will not further pursue the analysis of the electroweak sector in this work, as we are interested in the phenomenology of the coloured resonances, which are more abundantly produced at hadron colliders.

\subsection{Axial masses and scalar mixing}

From the $d_1^2$ term, we obtain a mass for the axial vectors:
\begin{align}
	\frac{f_1^2}{2} \Tr d_{1,\mu} d_1^{\mu} \supset \frac{f_1^2}{2} \Tr( \tilde g^2 \boldsymbol {\mathcal A}_\mu \boldsymbol {\mathcal A}^\mu - 2 \tilde g \partial_\mu \boldsymbol \pi_1 \boldsymbol {\mathcal A}^\mu  )\,,
\end{align}
while a mixing with $\pi_0$ is generated by the $d_{0,\mu} d_1^\mu$ term.
The mixing terms can be removed with an appropriate choice of gauge fixing, leaving a common mass term for all the axial vectors:
\begin{align}
	M_{\mathcal A} = \frac{\tilde g f_1}{\sqrt 2}\,.
\end{align}

The mesons $\pi_0$ and $\pi_1$ undergo a non-trivial mixing, analogous to the case studied in~\cite{BuarqueFranzosi:2016ooy}, hence we will simply recall the basics here. 
As the $d_i$ forms give
\begin{align}
	d_{i,\mu} = -\frac{\sqrt 2}{f_i} D_\mu \boldsymbol{\pi}_i + \cdots\,,
\end{align}
at leading order in the expansion, the Lagrangian contains a kinetic mixing of the form:
\begin{align}
	\mathcal L \supset \Tr( D_\mu \boldsymbol \pi_0 D^ \mu \boldsymbol \pi_0 + D_\mu \boldsymbol \pi_1 D^ \mu \boldsymbol \pi_1 + 2r\frac{f_1}{f_0} \, D_\mu \boldsymbol \pi_0 D^\mu \boldsymbol  \pi_1)\,.
\end{align}
Hence, one can define decoupled and canonically normalised fields $\pi_A$ and $\pi_B$ as 
\begin{align}
	\pi_0 &= \frac{\pi_A}{\sqrt{2} \sqrt{1+r\,f_1/f_0}} - \frac{\pi_B}{\sqrt{2} \sqrt{1-r\,f_1/f_0}}\,, \\
	\pi_1 &= \frac{\pi_A}{\sqrt{2} \sqrt{1+r\,f_1/f_0}} + \frac{\pi_B}{\sqrt{2} \sqrt{1-r\,f_1/f_0}}\,.
\end{align}
A linear combination of these states is eaten by the $\mathcal A_\mu$. The physical $\pi_P$ and unphysical $\pi_U$ states are given by
\begin{align}
	\pi_A=\cos\alpha \,\pi_P - \sin\alpha \,\pi_U, \qquad \pi_B = \sin\alpha\, \pi_P + \cos\alpha\, \pi_U
\end{align}
where
\begin{equation}
	\tan\alpha = - \sqrt{\frac{1+r\, f_1/f_0}{1-r\,f_1/f_0}}.
\end{equation}
Combining the above redefinitions yields
\begin{equation}\label{eq:piredefinition}
	\pi_0 = \pi_P \, \frac{1}{\sqrt{1-r^2 f_1^2/f_0^2}}, \qquad \pi_1 = \pi_U - \pi_P\, \frac{rf_1/f_0}{\sqrt{1- r^2 f_1^2/f_0^2}}.
\end{equation}
In the unitary gauge, only the $\pi_P$ remain in the spectrum, and they correspond to the pNGBs from the coset $\SU(6)/\SO(6)$.

\subsection{Decay channels}

We are now ready to determine the main decay modes for the heavy spin-1 resonances. They are generated by three types of interactions:
\begin{itemize}
    \item Couplings to pNGBs from the chiral Lagrangian in the strong sector, Eq.~\eqref{eq:Lagr};
    \item Couplings to quarks via the mixing of the colour octet to gluons;
    \item Partial compositeness couplings to top and bottom quarks.
\end{itemize}

The first type stems directly from the pNGB embedding in the effective Lagrangian.
We recall that, in the unitary gauge, the relevant terms simplify to
\begin{align}
	\mathcal L \supset\, & \frac{f_0^2}{2} \Tr d_{0,\mu} d_0^{\mu} + \frac{f_1^2}{2} \Tr d_{1,\mu} d_1^{\mu} + rf_1^2\, \Tr d_{0,\mu} d_1^{\mu} \nonumber \\
	&+ \frac{f^2_K}{2} \Tr e_{0,\mu} e_0^{\mu} + \frac{f^2_K}{2} \Tr e_{1,\mu} e_1^{\mu} - f^2_K \Tr e_{0,\mu} e_1^{\mu}\,.
\end{align}
We are interested in terms linear in the vector fields $\mathcal V_\mu/\mathcal A_\mu$ and with the smallest number of pNGBs.
It turns out that these only come as two independent traces:
\begin{align}
	\mathcal O_V &= i \Tr([\boldsymbol \pi, \partial_\mu \boldsymbol \pi] \boldsymbol{ V}^\mu),\\
	\mathcal O_{\mathcal A} &= \Tr( [\boldsymbol \pi, [\boldsymbol \pi, \partial_\mu \boldsymbol \pi]] \boldsymbol {\mathcal A}^\mu ) ,
\end{align}
where $V=\mathcal V,G,B$ is a generic vector. We recover explicitly that vectors couple to two pNGBs, while axial resonances can only couple to three pNGBs.
Both $\mathcal O_V$ and $\mathcal O_{\mathcal A}$ are hermitian.
After transforming the pions and vectors to the physical fields, we find that these operators come with coefficients 
\begin{align}
	C_{\mathcal V_{3}} &=  \frac{\tilde g (r^2-1) f_K^2}{f_0^2 \ (1-R^2)} \equiv g_{\rho\pi\pi}\label{eq:cv36}\,,\\
	C_{\mathcal V_8} &= \frac{(r^2-1) f_{K}^{2} }{f_{0}^{2} \ \left(1 - R^{2}\right)} (\tilde g \cos \beta_8 + \hat{g}_s \sin \beta_8) + 2\frac{1+R^2}{1-R^2} \hat{g}_s \sin \beta_8 \nonumber\\
 &= \frac{g_{\rho\pi\pi}}{\cos\beta_8} + 2\frac{1+R^2}{1-R^2} g_s \tan\beta_8\,, \label{eq:cv8} \\
	C_{\mathcal A} &= \frac{\sqrt 2 \tilde g r}{3} (1-r^2) (f_1^2-3f_K^2) \equiv g_{a3\pi}\,;
\end{align}
where $R = r f_1/f_0$. 
The details of this calculation are presented in \cref{app:decays}. The colour structure of the couplings among the various components are determined uniquely by the above traces in the $\SU(6)$ space.

The second type of couplings originates from the mixing of the gluon with $\mathcal V_8$, hence yielding a universal coupling of the massive resonance to quarks:
\begin{align}\label{eq:couplingV8qq}
	\mathcal L_\mathrm{fermions} \supset i\bar q  \slashed D q \supset \hat g_s\, \bar q\, \slashed G^a t^a_{\mathbf 3} \,q
    \to -g_s \tan\beta_8\, \bar q\, \slashed {\mathcal V}_8^a t^a_{\mathbf 3} \,q \equiv C_{\mathcal V_8}^{qq} \mathcal O_{\mathcal V_8}^{qq}\,,
\end{align}
where $t_{\mathbf 3}^a = \lambda^a/2$ are the colour generators for the fundamental irrep.
Note that the massless octet inherits a coupling $\hat{g}_s \cos \beta_8 \equiv g_s$, hence consistent with QCD gauge invariance.
A coupling to two gluons, instead, is not generated, as shown in \cref{app:gluons}.

Finally, the third type is generated by the coupling of the spin-1 resonances to the baryons \cite{Erkol:2006sa,Aliev:2009ei} that mix to top quarks via the partial compositeness mechanism. While the couplings generated by the strong dynamics are inherently vector-like, the chiral mixing of the physical states generates chiral couplings to the mass eigenstates. Details of the origin of these couplings are presented in \cref{app:PC}.
Such couplings always exist for the colour octet states, and they can be parameterised as
\begin{align}\label{eq:LagPC}
    \mathcal{L}_\text{PC} \supset &\; \bar{t} \slashed {\mathcal V}_8^a t^a_{\mathbf 3} \left( g_{\rho t t, LL} P_L + g_{\rho t t, RR} P_R \right) t +  \bar{b} \slashed {\mathcal V}_8^a t^a_{\mathbf 3} \left( g_{\rho b b, LL} P_L  \right) b +\nonumber \\
    &  \bar{t} \slashed {\mathcal A}_8^a t^a_{\mathbf 3} \left( - g_{a t t, LL} P_L + g_{a t t, RR} P_R \right) t +  \bar{b} \slashed {\mathcal A}_8^a t^a_{\mathbf 3} \left( - g_{a b b, LL} P_L \right) b \,,
\end{align}
where $P_{L,R}$ are chiral projectors and we only consider the electric part of the coupling. In the models under consideration, we have that $g_{\rho/a b b, LL} \simeq  g_{\rho/a t t, LL}$ while the bottom coupling is only left-handed at leading order. Note that all the above couplings are of order $\tilde{g}$, while the chiralities are distinguished by the different mixing angles from partial compositeness.
The non-octet resonances, $\mathcal{V}_3$ and $\mathcal{A}_6$, couple to a pair of quarks via partial compositeness only in models where the two resonances have baryon number $2/3$ and charge $\pm 4/3$, hence leading to two top decay channels. The effective couplings can be parameterised as 
\begin{equation}
\mathcal{L}_\text{PC} \supset g_{\rho t t, LR}\ \overline{t^c} \slashed{\mathcal{V}}_3 t +  g_{a t t, LR}\ \overline{t} \slashed{\mathcal{A}}_6 t^c + \text{h.c.} 
\end{equation}
where the superscript $c$ indicates charge conjugation. As the currents contain effectively one left-handed and one right-handed top, the couplings must be suppressed by the EW scale $v$ divided by the Higgs decay constant as compared to the octet couplings.

Note finally that, while the first two types of couplings are completely determined by the chiral Lagrangian in Eq.~\eqref{eq:Lagr}, the third one is more model dependent.
In fact, the value of the couplings depend on the quantum numbers of baryons that mix with the elementary top fields, and on the value of the mixing angles.~\footnote{Couplings to light quarks could also be generated by partial compositeness, however their couplings will be generically suppressed by the small mixing required by the lightness of the quark masses. Hence, such contributions can be neglected compared to the mixing with the gluon.} Hence, they cannot be predicted in a model-independent way and we will leave them as free parameters.

\subsection{Independent parameters}
The effective Lagrangian for the coloured spin-1 resonances contains six free parameters: $f_0$, $f_1$, $f_K$, $r$, $\hat{g}_s$, and $\tilde g$ (the mixing angles depend only on $\tilde g$). 
As we have seen, $\hat{g}_s$ can be fixed by the physical coupling of the massless gluons, as in Eq.~\eqref{eq:gsphys}.
We can trade $f_1$ and $f_K$ for masses:
\begin{align}
	f_1 = \frac{\sqrt 2 M_{\mathcal A}}{\tilde g}, \qquad f_K = \frac{\sqrt 2 M_{\mathcal V_3}}{\tilde g} = \frac{\sqrt 2 M_{\mathcal V_8}}{ \sqrt{\tilde g^2 + \hat g_s^2} } = \frac{\sqrt 2 M_{\mathcal V_1}}{ \sqrt{\tilde g^2 + \hat g'{}^2} }\,,
\end{align}
hence we can choose as input parameters $M_{\mathcal V_8}$ and the ratio
\begin{align}
    \xi = \frac{M_{\mathcal A}}{M_{\mathcal V_8}}\,.
\end{align}
Note that the relation between the two vector masses only depends on the octet mixing angle, i.e.\ on $\tilde g$, as $M_{\mathcal{V}_3} = M_{\mathcal{V}_8} \cos \beta_8$.
We can further use as an input the physical decay constant of the pNGBs $f_\chi$, which enters the couplings of the physical $\pi_P$ states and reads:
\begin{align}\label{eq:fchi}
	f_\chi = \sqrt{f_0^2 - r^2 f_1^2}\,.
\end{align}
Finally, another input parameter can be the coupling of the vectors to the pion, $g_{\rho\pi\pi}$, which can be measured on the lattice, for instance. It relates to the Lagrangian parameters as follows:
\begin{align}\label{eq:grpp}
	g_{\rho\pi\pi} = \left. C_{\mathcal V_8}\right|_{\beta_8\to 0} = C_{\mathcal V_{3/6}} = \frac{\tilde g (r^2-1) f_K^2}{f_0^2 \ (1-R^2)}\,.
\end{align}
Solving \cref{eq:fchi,eq:grpp} for $f_0$ and $r$ yields
\begin{align}
	f_0 = \sqrt{ \frac{f_1^2 f_\chi^2}{f_K^2} \frac{g_{\rho\pi\pi}}{\tilde g} + f_1^2 + f_\chi^2 }\,, \qquad r = \sqrt{1 + \frac{f_\chi^2 g_{\rho\pi\pi}}{f_K^2 \tilde g}}\,.
\end{align}

In summary, this leaves us with five independent input parameters:
\begin{align}
	\tilde g, \quad g_{\rho\pi\pi}, \quad M_{\mathcal V_8}, \quad \xi, \quad f_\chi\,.
\end{align}
As already mentioned, in addition we have the couplings to top and bottom quarks generated by top partial compositeness.

\subsection{Generalisation to $\SU(6)/\Sp(6)$ and $\SU(3)\times\SU(3)/\SU(3)$}

For the $\SU(6)/\Sp(6)$ case, the computation of the effective Lagrangian follows the same patterns as described above, with the only difference in the broken and unbroken generators. Effectively, this implies that colour charges of the non-octet states are interchanged: $\mathcal{V}_6$ and $\mathcal{A}_3$ (as well as $\pi_3$). The coefficients of the various couplings and mass values, however, follow the same results as above.

For the case $\SU(3)\times \SU(3)/\SU(3)$, the action of the symmetries are slightly different in form. However, for this coset all vector and axial resonances, as well as the pNGBs, transform as octets. Hence, the effective interactions are the same as above, once the non-octet states are removed.

\section{Phenomenology}\label{sec:pheno}

\begin{table}[htb]
\centering\small
\begin{tabular}{|c|c|c||c|c|c|c|c|}
\hline
& Models & $\chi$ ($R,Y,B$) & $\pi$ & $\mathcal{V}^\mu$ & $\mathcal{A}^\mu$ & $\Psi$ &di-quark\\ \hline\hline
C1 & M1-2 & (R,$-\frac{1}{3}$,$\frac{1}{6}$) & $8_0,\, {\color{red} 6_{-2/3}}$ & $8_0,\, 1_0,\, {\color{red} 3_{2/3}}$ & $8_0,\, {\color{red} 6_{-2/3}}$ & $8,\, 1,\, {\color{red} 3},\, {\color{red} 6}$ & none\\ \hline
C2 & M3-4, M8-11 & (R,$\frac{2}{3}$,$\frac{1}{3}$) & $8_0,\, {\color{blue} 6_{4/3}}$ & $8_0,\, 1_0,\, {\color{blue} 3_{-4/3}}$ & $8_0,\, {\color{blue} 6_{4/3}}$ & ${\color{red} 3}$ & $\pi_6,\mathcal{V}^\mu_3,\mathcal{A}^\mu_6$\\ \hline
C3 & M5 & (Pr,$-\frac{1}{3}$,$\frac{1}{6}$) & $8_0,\, {\color{red} 3_{2/3}}$ & $8_0,\,  1_0,\, {\color{red} 6_{-2/3}}$ & $8_0,\, {\color{red} 3_{2/3}}$ & $8,\, 1,\, {\color{red} 3},\, {\color{red} 6}$ & none\\ \hline
C4 & M6-7 & (C,$-\frac{1}{3}$,$\frac{1}{6}$) & $8_0$ & $8_0,\,  1_0$ & $8_0$ & $8,\, 1,\, {\color{red} 3},\, {\color{red} 6}$ & none\\ \hline
C5 & M12 & (C,$\frac{2}{3}$,$\frac{1}{3}$) & $8_0$ & $8_0,\,  1_0$ & $8_0$ & ${\color{red} 3}$ & none\\ \hline
\end{tabular}
\caption{\label{tab:QCDres} Properties of the spin-0 ($\pi$), spin-1 ($\mathcal{V}^\mu$, $\mathcal{A}^\mu$) and spin-1/2 ($\Psi$) lightest resonances in the 12 models, grouped in 5 classes. Each class is determined by the properties of the $\chi$ species, listed in the second column by irrep type (R for real, Pr for pseudo-real and C for complex). For the resonances, the colours indicate the baryon numbers, with black for $B=0$, red for $B=\pm 1/3$ and blue for $B=\pm 2/3$. In the last column we indicate the bosons that can decay into a di-quark state ($tt$). }
\end{table}

The twelve models under consideration allow us to predict the quantum numbers of the lightest coloured resonances. 
Following the properties of the fermion species $\chi$, they can be grouped into five classes, as shown in \cref{tab:QCDres}. 
For the fermionic states, the electroweak charges depend on the configuration of the $\psi$ fermions inside the chimera baryons, and a full classification is possible, but beyond our purposes. 
In fact, we will assume here a lattice and QCD inspired mass hierarchy, where the baryon-like states are heavier than the spin-1 states, which are heavier than the pNGBs.
Henceforth, the heavy baryons do not have a direct relevance for the phenomenology of the spin-1 states, except for the fact that their couplings  can generate a direct coupling of the spin-1 resonances to a pair of tops via the top partial compositeness mixing, as discussed in the previous section.

The coloured spin-1 resonances, therefore, can be produced via their QCD interactions at hadron colliders. This leads to pair production for all types of states. The only one that also features single-production is the vector colour octet, as it inherits a universal coupling to all quarks via its mixing to the gluons. As the masses of the spin-1 resonances are expected to be of the same order, we will first study the LHC limits on the vector colour octet to determine the smallest allowed mass. Before doing that, however, it is important to recall the properties of the coloured pNGBs, which appear in the decays of all spin-1 resonances. Finally we will present first results for future high energy hadron colliders, which could access pair production of all the resonances.

\subsection{Coloured pNGB decays}

The phenomenology of the coloured pNGBs have been studied in several works and contexts \cite{Cacciapaglia:2015eqa,Belyaev:2016ftv,Cacciapaglia:2020vyf}, hence we will here only remind their main features.

A colour octet pNGB is ubiquitous to all models. It always features two types of couplings: a coupling to gauge bosons generated by a topological anomaly and one to tops generated by partial compositeness \cite{Belyaev:2016ftv,Cacciapaglia:2020vyf}. Which one dominates, however, depends crucially on the details of the model, as their origin is rather different in nature. Note that the anomaly dominantly consists of couplings to two gluons, however it also generates suppressed couplings to $g\gamma$ and $gZ$, which provide interesting and clean final states \cite{Belyaev:1999xe,Cacciapaglia:2020vyf}. Nevertheless, to simplify the analysis and focus on existing searches, we will neglect the single-gluon decay channels in the following. 

The decays of the non-octets depend crucially on the scenario at hand. Following the classification in Table~\ref{tab:QCDres}, we distinguish four cases:
\begin{alignat}{2}
	&\text{C1}: && \qquad \pi_8 \to t\bar t,\, gg ; \,\, \pi_6 \to bb \,,\\
	&\text{C2}: && \qquad \pi_8 \to t\bar t,\, gg ; \,\, \pi_6 \to tt\,, \\
	&\text{C3}: && \qquad \pi_8 \to t\bar t,\, gg ; \,\, \pi_3 \to \bar b\bar s \text{ or } t\bar \nu,\, b\tau^+\,,\\
	&\text{C4-5}: && \qquad \pi_8 \to t\bar t,\, gg\,.
\end{alignat}
In C2, the sextet has baryon number $2/3$ and charge $4/3$, hence partial compositeness will generate an unsuppressed coupling to two right-handed tops \cite{Cacciapaglia:2015eqa}. In C1 and C3, the sextet and triplet have baryon number $\mp 1/3$ and charges $\mp 2/3$, respectively, hence they are not allowed to decay into standard model fermions by partial compositeness alone. Their decays must, therefore, be generated by specific operators that need to violate either baryon or lepton number. Considering the standard model gauge quantum numbers \cite{Carpenter:2021rkl}, the allowed final states are listed above. The di-quark final state violates baryon number by one unit, $\Delta B=1$, and we consider preferential couplings to heavier flavours (while this is not strictly required). For the triplet, decays to a quark and a lepton can be envisioned, violating lepton number by one unit, $\Delta L=1$. They can be generated in some models by partial compositeness extended to leptons \cite{Cacciapaglia:2021uqh}, hence naturally involving the third (heavier) family. In such case, the triplet effectively behaves like a composite leptoquark \cite{Gripaios:2009dq}. We remark that the $B$ or $L$ violating couplings can be rather small, however they provide the only decay channel for the sextet and triplet states. Depending on the value of these couplings, therefore, they decay promptly, as we consider in the following, or lead to displaced vertices and anomalously massive hadronic tracks \cite{Kraan:2004tz}.

The ubiquitous colour octet can be searched at the LHC via QCD pair production: in the following, we will assume dominant top couplings, hence leading to a four tops final state. A recent reinterpretation \cite{Darme:2021gtt} of a CMS search \cite{CMS:2019rvj} leads to a conservative lower bound of $1.25$~TeV for the colour octet mass. Note that in C2 models, the contribution of the sextet can further push up this limit. 
Dedicated searches also exist for the leptoquark decays of the triplet in C3 models, where both $b\tau$ and $t\nu$ final states have been searched for by ATLAS and CMS \cite{ATLAS:2021jyv,CMS:2020wzx,CMS:2023qdw,ATLAS:2023uox,ATLAS:2024huc} yielding bounds between $1.25$ and $1.46$~TeV, depending on the branching ratios in the two channels. The bounds on this mass
are significantly lower of about $770$~GeV if $\pi_3$ decays dominantly
into light quarks \cite{ATLAS:2017jnp,CMS:2022usq}.

\subsection{Colour octet single production at the LHC}

\begin{figure}
    \centering
    \includegraphics{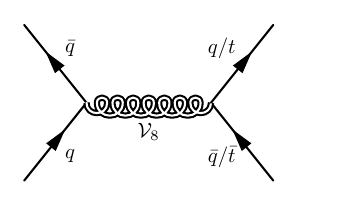}
    \includegraphics{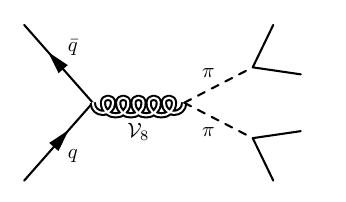}
    \caption{Feynman diagrams of $\mathcal V_8$ single production and decay into quarks or pNGBs.}
    \label{fig:feynmanv8single}
\end{figure}

\begin{figure}
    \centering
    \includegraphics[width=\linewidth]{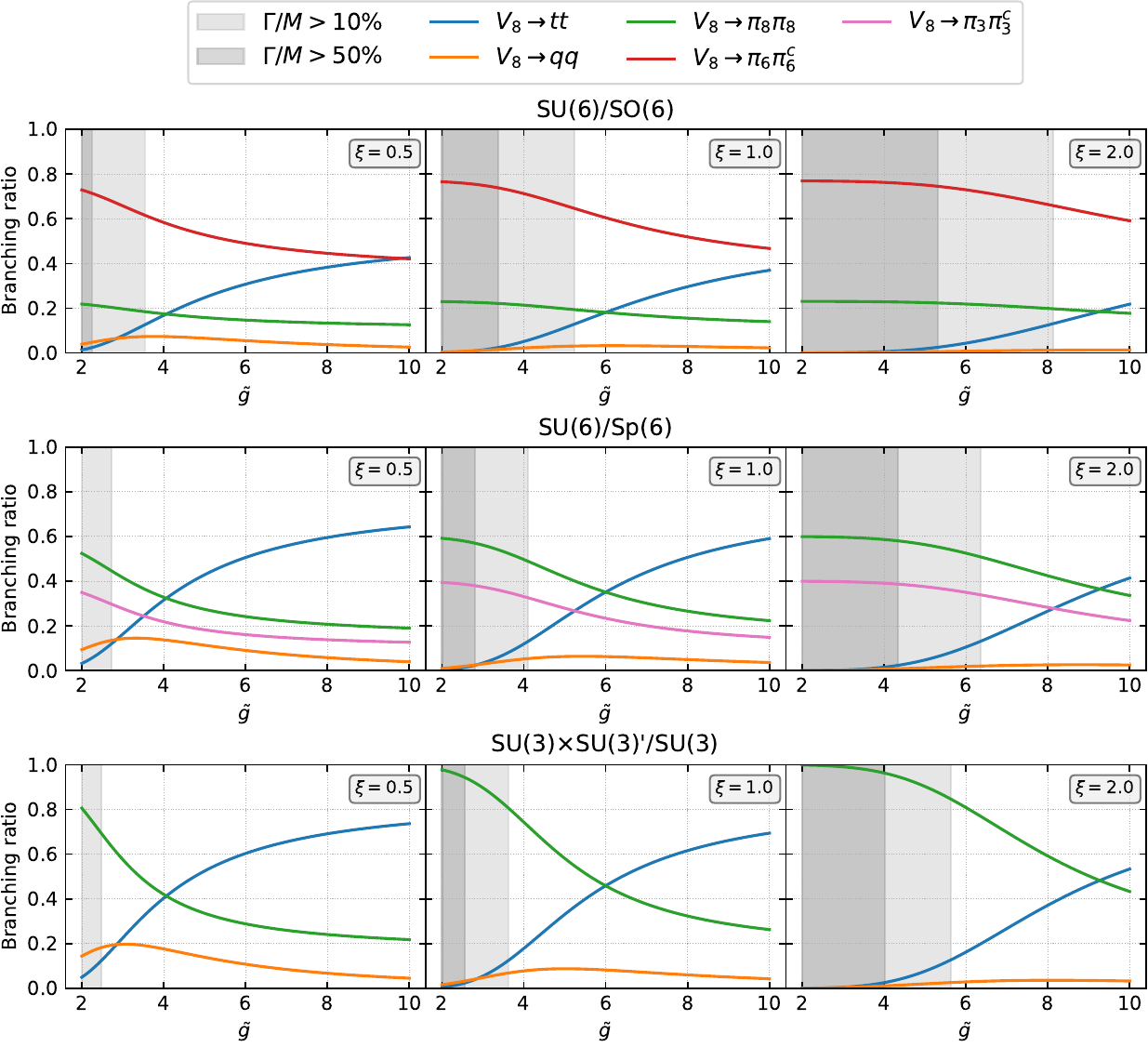}
    \caption{Sample branching ratios for $g_{\rho \pi \pi}=1$ and $M_{\mathcal V_8}=4.5$~TeV. We also fix $m_\pi = 1.4$~TeV, $f_\chi=1$~TeV and the coupling to top quarks to 1. Note that here $q=u,d,s,c,b$.}
    \label{fig:branchingratiosmain}
\end{figure}

Via mixing to the gluons, the vector octet $\mathcal V_8$ inherits a coupling to all quarks, see \cref{eq:couplingV8qq}, allowing it to be singly produced. 
This coupling is suppressed by a mixing angle that depends only on $\tilde{g}$, hence it is reduced for large $\tilde{g}$, see also Eq.~\eqref{eq:gsphys}.
Nevertheless, even for moderate values, the single production cross section dominates over pair production.
Henceforth, the colour octet is the main resonance to be hunted at hadron colliders, and it has been considered in the literature in various composite contexts \cite{Lillie:2007yh}. Typically, decays into two quarks are considered, while we will also include decays into two coloured pNGBs
as shown in Fig.~\ref{fig:feynmanv8single}.

In the models under consideration, the possible decay modes of the vector octet can be classified as follows:
\begin{alignat}{2}
	&\text{C1-2}: \qquad &&\mathcal V_8 \to q\bar q,\; b\bar b,\; t\bar t,\; \pi_8 \pi_8,\; \pi_{6} \pi_{6}^c,\\
	&\text{C3}: &&\mathcal V_8 \to q\bar q,\; b\bar b,\; t\bar t,\; \pi_8 \pi_8,\; \pi_{3} \pi_{3}^c, \\
	&\text{C4-5}: &&\mathcal V_8 \to q\bar q,\; b\bar b,\; t\bar t,\; \pi_8 \pi_8,
\end{alignat}
where C1 and C2 are distinguished by the decays of the sextet pNGB.
The decays into light quarks $q=u,d,c,s$ feature flavour-independent branching ratios, while bottom and top quark channels receive additional contributions from partial compositeness, see \cref{eq:LagPC}, leading to different branching ratios. 
Finally, the relative strength of the pNGB channels is determined purely by colour factors, assuming their masses are equal, and we find
\begin{equation}\label{eq:pngb_brs}
    \frac{\text{BR} (\pi_6 \pi_6^c)}{\text{BR} (\pi_8 \pi_8)} = \frac{10}{3}\,, \qquad \frac{\text{BR} (\pi_3 \pi_3^c)}{\text{BR} (\pi_8 \pi_8)} = \frac{2}{3}\,.
\end{equation}
The importance of each channel depends on the parameter space, and we provide some benchmarks in Fig.~\ref{fig:branchingratiosmain}. 
The relevance of the decays into light quarks and pNGBs depends mainly on the $\tilde{g}$ and $g_{\rho\pi\pi}$ couplings. On the one hand, the partial width to light quarks is controlled by the mixing angle to gluons and it decreases for increasing $\tilde{g}$. On the other hand, the partial width to pNGBs receives a dominant contribution proportional to $g_{\rho\pi\pi}$: the dependence on $\tilde{g}$ is such that this partial width also decreases for increasing $\tilde{g}$. For very small $\tilde{g}$, instead, the second term in Eq.~\eqref{eq:cv8} starts becoming relevant, thus explaining the drop in the $qq$ branching ratio observed in Fig.~\ref{fig:branchingratiosmain}. The scaling in $\tilde{g}$ also explains why the total width of $\mathcal{V}_8$ increases for small values and for large octet masses. Finally, the branching ratio to top (and bottom) receives a dominant contribution from partial compositeness, which do not scale with $\tilde{g}$ and hence dominates for large values.

\begin{figure}
	\centering
	\includegraphics[width=0.5\linewidth]{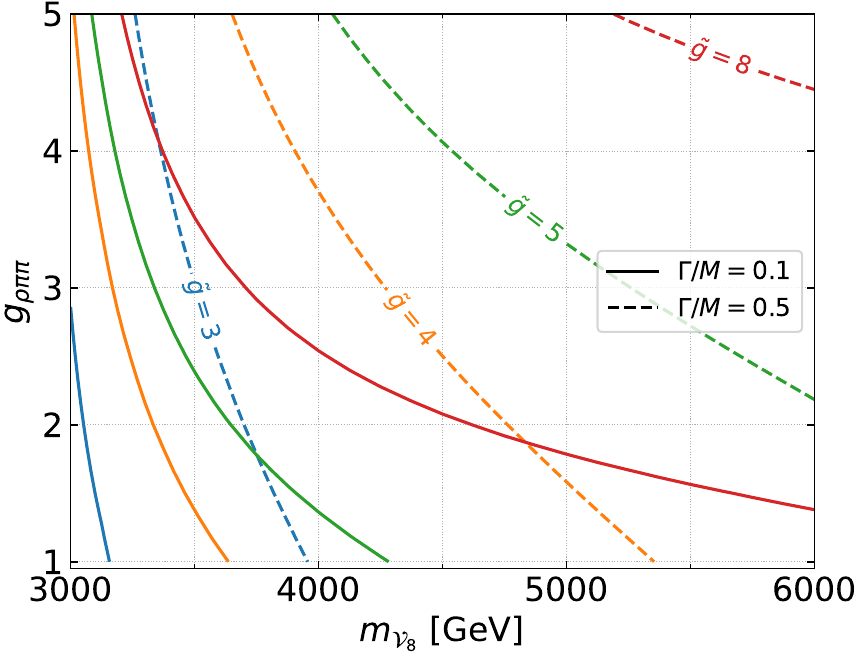}
	\caption{Isocurves of width/mass for $\xi=1.0$ and $f_\chi=1$~TeV for different values of $\tilde{g}$ for classes C1-2. The coupling to top quarks is fixed to 1.}
	\label{fig:isoWoverM}
\end{figure}

One caveat is that the coupling $g_{\rho\pi\pi}$ and the coupling to baryons are expected to be large in the strong theory, hence the colour octet will tend to have large width as compared to its mass. 
To quantify this important effect, we show in Fig.~\ref{fig:isoWoverM} curves of fixed width over mass ratios for different values of $\tilde{g}$ as a function of $g_{\rho\pi\pi}$ and the octet mass. The plots highlight the fact that the width can be larger than 50\% of the mass, especially for small values of $\tilde{g}$, hence invalidating the treatment of the octet in a standard narrow width approximation. 

\begin{figure}
	\centering
	\begin{subfigure}{0.48\linewidth}
	   \includegraphics[width=\linewidth]{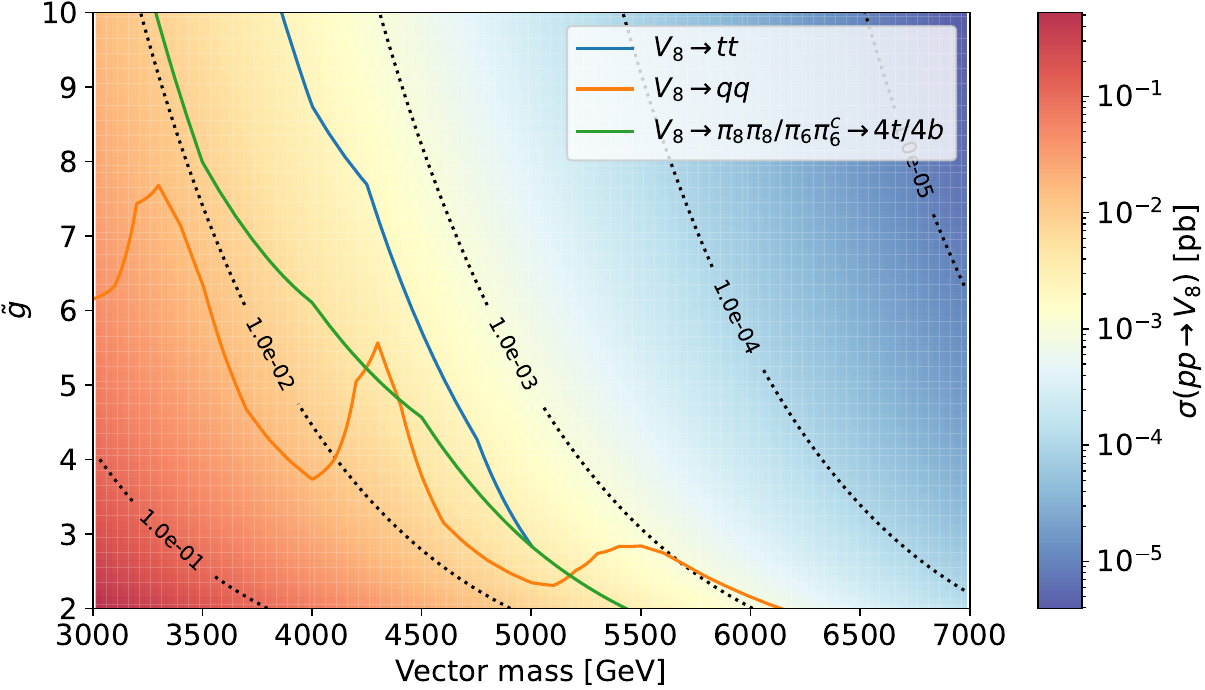}     
        \caption{Class C1}
	\end{subfigure}
    \begin{subfigure}{0.48\linewidth}
	   \includegraphics[width=\linewidth]{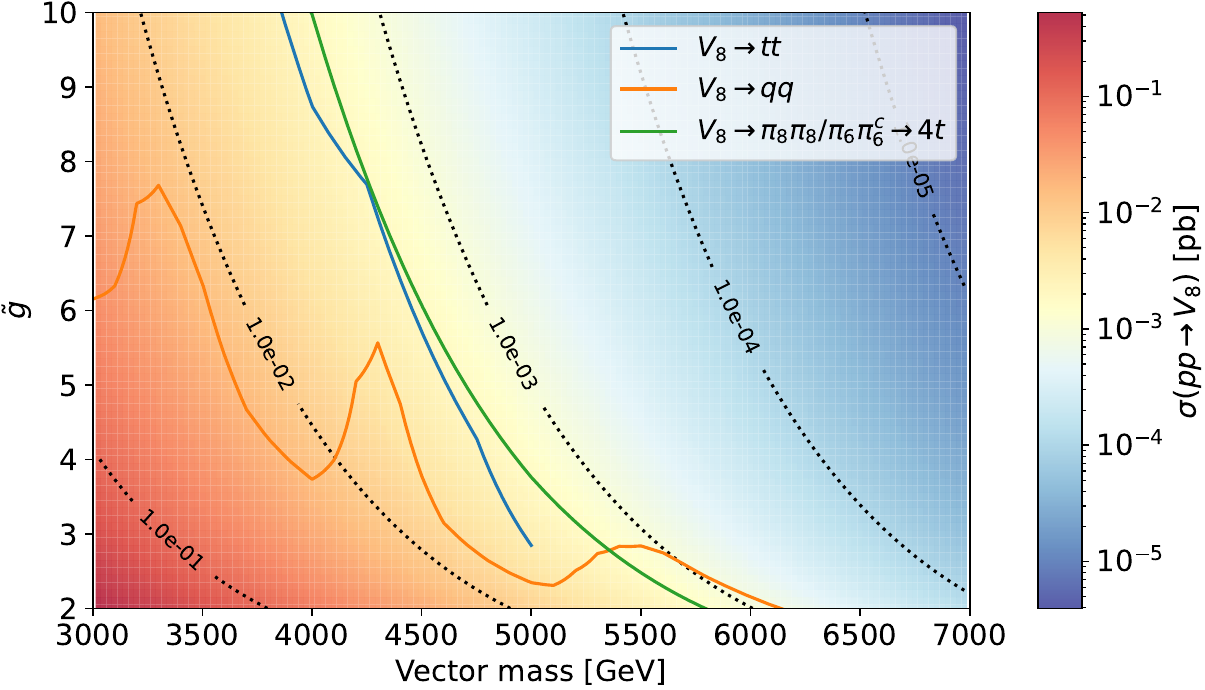}     
    \caption{Class C2}
	\end{subfigure}

    \begin{subfigure}{0.48\linewidth}
	   \includegraphics[width=\linewidth]{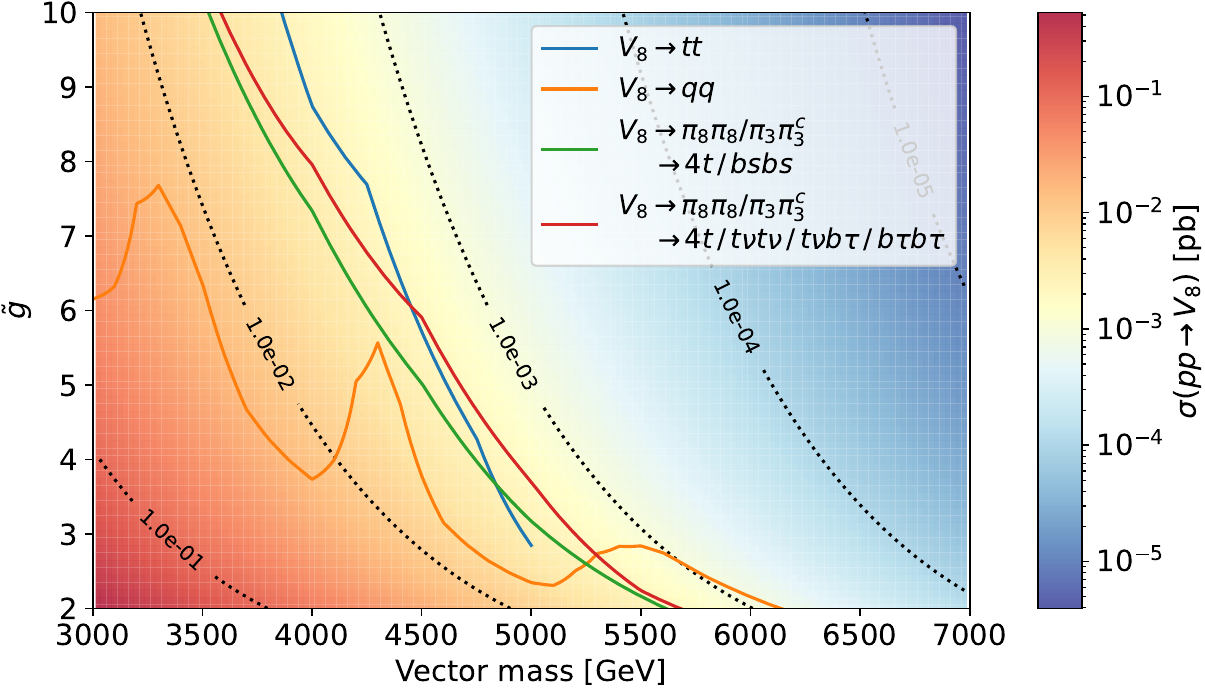}     
        \caption{Class C3}
	\end{subfigure}
     \begin{subfigure}{0.48\linewidth}
	   \includegraphics[width=\linewidth]{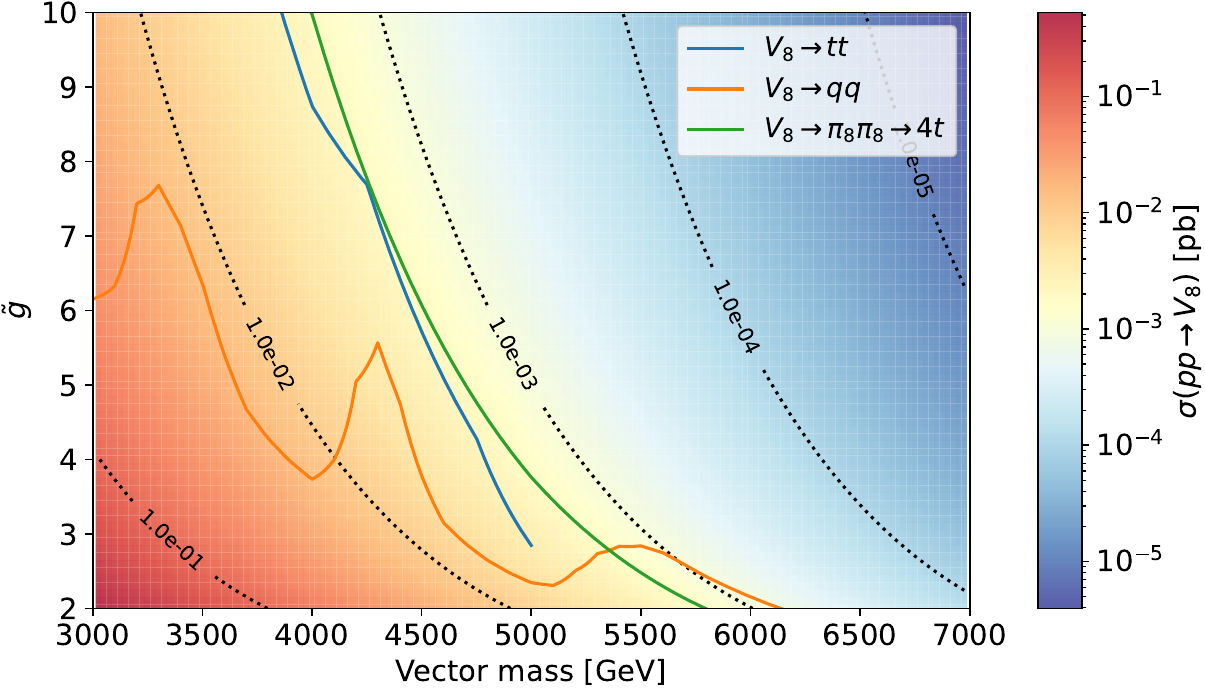}     
        \caption{Classes C4-5}
	\end{subfigure}
    
	\caption{\label{fig:massbounds} Bounds on vector octet single production for the model classes defined in \cref{tab:QCDres}. The heat map and the dotted contours indicate the single production cross section. The region to the left and below the coloured lines is excluded. The bounds are determined assuming 100\% branching ratio into the indicated channel. For the decays in pNGBs, the branching ratios in \cref{eq:pngb_brs} are taken into account. }
\end{figure}

Finally, the current bounds on the colour octet mass crucially depend on the branching ratios in the three channels: light quarks, pNGBs and tops, as they are controlled by different couplings. As a model-independent estimate of the bounds we, therefore, decided to show limits assuming 100\% branching ratios in the three channels, as shown in Fig.~\ref{fig:massbounds} for the four nonequivalent classes C1, C2, C3 and C4-5. The lines are extracted from the following searches:
\begin{itemize}
    \item di-jet: Search for high mass di-quark resonances \cite{CMS:2019gwf};
    \item di-top: Search for $t\bar t$ resonances \cite{ATLAS:2020lks};
    \item pNGBs: Recasts of SUSY searches \cite{ATLAS:2021twp,ATLAS:2019fag,ATLAS:2022ihe} implemented in \texttt{CheckMATE}\footnote{
For the simulation of signal events, we implemented the relevant interactions as a \texttt{FeynRules} \cite{Alloul:2013bka} model at leading order. For each mass point, we generate $10^4$ events using \texttt{MadGraph5\_aMC@NLO} \cite{Alwall:2014hca} version 3.5.3, in association with the parton densities in the \texttt{NNPDF 2.3} set \cite{Ball:2012cx,Buckley:2014ana}.
We then interfaced the events with \texttt{Pythia8} \cite{Sjostrand:2014zea} for showering and hadronisation. 
The resulting showered signal events are analysed with  \texttt{CheckMATE} \cite{Drees:2013wra,Dercks:2016npn} (commit number \texttt{1cb3f7}). To this end, events are reconstructed using \texttt{Delphes 3} \cite{deFavereau:2013fsa} and the anti-$k_T$ algorithm \cite{Cacciari:2008gp} implemented in \texttt{FastJet} \cite{Cacciari:2011ma}. 
We also ran the events against the searches and SM measurements implemented in \texttt{MadAnalysis5} \cite{Conte:2012fm,Conte:2014zja,Dumont:2014tja,Conte:2018vmg} version 1.10.9beta and \texttt{Rivet} \cite{Bierlich:2019rhm} version 3.1.8 in combination with \texttt{Contur} \cite{Butterworth:2019wnt, Buckley:2021neu} version 2.2.4, but both yielded subdominant bounds compared to \texttt{CheckMATE}.}.
\end{itemize}
The coloured heatmap indicates the Drell-Yan cross section, which only depends on $\tilde{g}$ and the octet mass, while the region below and to the left of the lines is excluded. The results show that the mass limits are roughly the same for all cases, and comparable for the three decay modes. Hence, the mass limits are in the range of $4$ to $5$~TeV. Note that the region for small $\tilde{g} \lesssim 3$ cannot be trusted as it corresponds to widths above 50\% of the mass, c.f. Fig.~\ref{fig:isoWoverM}. 

The High-Luminosity run at the LHC will certainly allow to further improve the mass limits on the vector octet, with the caveat that dedicated searches or reinterpretations will be needed to take into account the large width. However, even with the current bounds in Fig.~\ref{fig:massbounds}, we can infer that pair production of all spin-1 resonances will be very small at the LHC, hence making their detection unlikely. In the next section, therefore, we will discuss pair production at a future high energy hadron collider.

\subsection{Pair production at future high-energy hadron colliders}

\begin{figure}
	\centering
	\includegraphics[width=0.6\linewidth]{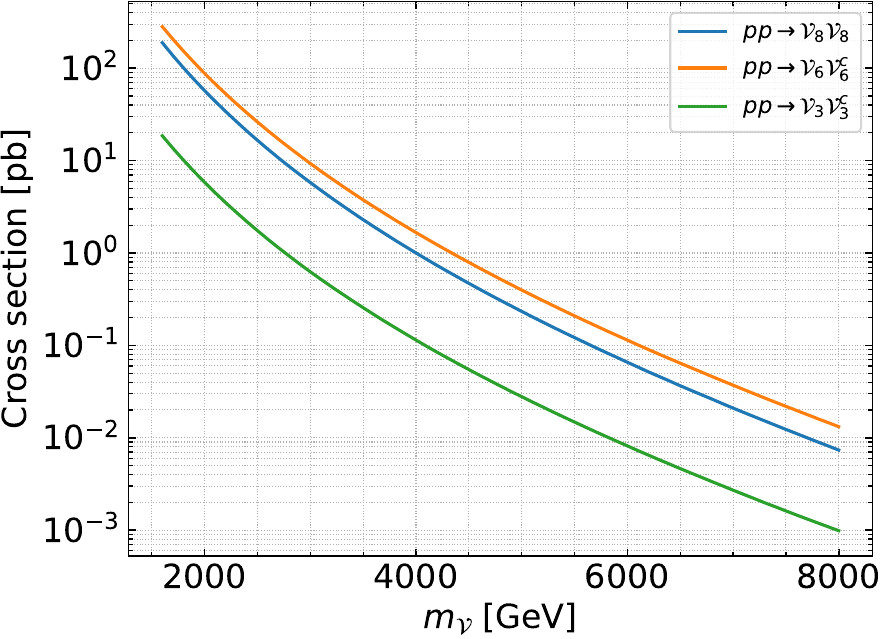}
	\caption{\label{fig:xsec} Pair production cross section at $\sqrt s = 100$~TeV for vectors in the sextet, octet and triplet representations. The same values apply for axial vectors.}
\end{figure}

Future hadron collider projects are expected to reach energies well above the LHC, with expectations up to $100$~TeV \cite{FCC:2018vvp,Tang:2022fzs}. At such energies, pair production of the vector and axial resonances will be accessible. In principle, single production of the vector octet remains the leading channel, with the caveat that at large masses the width will also increase and hence affect the search strategy. In this section, we will focus on pair production. The cross sections only depend on the QCD quantum numbers of the spin-1 states, and they are the same for vector and axial-vector states. In Fig.~\ref{fig:xsec} we show them as a function of the masses for $pp$ collisions at $100$~TeV centre of mass, using the \texttt{NNPDF 2.3} PDF set \cite{Ball:2012cx,Buckley:2014ana}. Hence, the most abundantly produced states will be the sextets, followed by octets, while triplet production is about one order of magnitude below.

In the following, we describe the main features to be expected from pair production, leaving a detailed analysis for future work. Firstly, we should stress that single production of the vector octet remains the leading discovery channel, including non-resonant effects that are relevant in the case of very large width. Hence, at a future $100$~TeV collider, one would expect to discover the octet before pair production becomes relevant.

The sextets, which feature the largest pair production cross sections, are present in the classes C1, C2 and C3. The decays can be classified as follows:
\begin{eqnarray}
        \mathcal{A}_6 &\to& \pi_8 \pi_8 \pi_6\ (t\bar{t}t\bar{t}bb)\; \text{and} \; \pi_6 \pi_6^c \pi_6 \ (\bar{b}\bar{b}bbbb)\quad \text{in C1}\,,\\
        \mathcal{A}_6 &\to& tt\; \text{or} \; \pi_8 \pi_8 \pi_6\ (t\bar{t}t\bar{t}tt)\; \text{and} \; \pi_6 \pi_6^c \pi_6 \ (\bar{t}\bar{t}tttt)\quad \text{in C2}\,,\\
        \mathcal{V}_6 &\to& \pi_8 \pi_3^c \to (t\bar{t}) (\bar{b}\bar{s}\; \text{or}\; q l) \quad \text{in C3}\,.
\end{eqnarray}
The two-body decay of $\mathcal{V}_6$ in C3 is driven by $g_{\rho\pi\pi}$, hence is will likely produce a large decay width. Instead, thanks to the three body final state for $\mathcal{A}_6$ in pNGBs, we expect the axial widths to remain small compared to the mass. In the cases C2, a competitive decay into two tops is also present: the two-body top decay width is, in fact, suppressed by $v^2/f^2 \leq 0.04$ (for $f \geq 1$~TeV), while the three-pNGB channel is suppressed by a phase space factor as compared to the two-body channel. Hence, we expect the two to lead to competitive branching ratios. In both cases C1 and C2, the pair production of the sextet will lead to final states with many top and bottom jets.

The axial colour octet $\mathcal{A}_8$ also has a sizeable pair cross section, and it also has a leading decay channel into two tops in all cases. Hence, it will generate 4-top final states with potentially large width effects.

Finally, the colour triplets, present in C1, C2 and C3, lead to the following decays:
\begin{eqnarray}
        \mathcal{V}_3 &\to& \pi_8 \pi_6^c\ (t\bar{t}\bar{b}\bar{b})\; \quad \text{in C1}\,,\\
        \mathcal{V}_3 &\to& \bar{t}\bar{t}\; \text{or} \; \pi_8 \pi_6^c\ (t\bar{t}\bar{t}\bar{t})\quad \text{in C2}\,,\\
        \mathcal{A}_3 &\to& \pi_8 \pi_8 \pi_3\; (t\bar{t}t\bar{t}+\bar{b}\bar{s}\; \text{or}\; q l)\; \text{or}\; \pi_3 \pi_3^c \pi_3 \; (\bar{b}\bar{s}bs\bar{b}\bar{s}\; \text{or}\; ql\bar{q}lql)\quad \text{in C3}\,.
\end{eqnarray}
For the vectors, the decays will be largely dominated by the $\pi_8\pi_6^c$ channels, as the di-top coupling in C2 is suppressed by $v/f$. The caveat remains that the vector widths may be large in most of the allowed parameter space. Instead, the axial in C3 remains narrow, leading to interesting final states rich in tops and possibly leptons, if the $\pi_3$ decays violate lepton number.

\section{Conclusions} \label{sec:conclu}

We have investigated the phenomenology of spin-1 resonances in Composite Higgs Models carrying QCD charges, with particular attention on production and decay modes, LHC bounds, and future hadron collider prospects. We have in particular focused on models which allow for fermionic UV completions \cite{Ferretti:2016upr,Belyaev:2016ftv} as they provide detailed information on the quantum numbers and properties of the bound states.
We have worked out their properties for three types of cosets, $\SU({6})/\SO({6})$, $\SU({6})/\Sp({6})$ and
$\SU({3}) \times \SU({3})/\SU({3})$ and the most relevant production and decay channels at present and future pp-colliders. The considered cosets are symmetric and they, therefore,  contain two sets of spin-1 resonances: vector states that couple to two pNGBs and axial-vector states that couple to three pNGBs.

In all scenarios, the vector $\mathcal{V}_8$ in the adjoint representations of colour $\SU({3})$ is present, and it 
mixes with the QCD gluon octet. Thanks to the mixing, this state can be singly produced at hadron colliders via Drell-Yan, whereas all other states can only be pair-produced. The $\mathcal{V}_8$ can either decay into a quark pair or into two pNGBs leading in all cases to the final states $q\bar{q}$ ($q\ne t$), $t\bar{t}$ and $4t$. In a coset with triplet or sextet pNGBs, in addition one has a subset of the following final states: $4b$, $2b2s$ or $2 t 2\nu$, $2 b 2 \tau$ and $t\nu b\tau$. We have investigated in all cases bounds on the mass ranging from $3.5$~TeV to $6$~TeV
from existing LHC data. We have focused on scenarios where $\mathcal{V}_8$ has a sufficiently small decay width so that the narrow width approximation holds. Hence, pair production is only relevant for a future high energy hadron collider, where pairs of the sextets and octets will be abundantly produced. We classified all permitted final states, which are typically rich in top quarks and leptons. Further studies, including the large width case and specific prospects for future 100 TeV pp-collider, as well as more model dependent signatures will be analysed in a separate publication.

\section*{Acknowledgements}

We thank Alan Cornell for discussions and collaboration during the initial stages of this project. We are grateful to the Mainz Institute for Theoretical Physics (MITP) of the DFG
Cluster of Excellence PRISMA+ (Project ID 39083149) for its hospitality and support
during the final stages of this work.
This work has been supported by the “DAAD, Frankreich” and “Partenariat Hubert Curien (PHC)” PROCOPE 2021-2023, project number 57561441.
M.K. and W.P. are supported by DFG, project 
no.~PO-1337/12-1.
M.K. is supported by the ``Studienstiftung des deutschen Volkes''.

\appendix

\section{Details on the calculation}

\subsection{Conventions}\label{app:conventions}
It is convenient to embed a field $\phi_{\mathbf r}$ in the $\mathbf r$ irrep of QCD within $3\times3$ matrices in the notation of \cite{Carpenter:2021rkl}:
\begin{align}
	\phi_3 = \phi_{3,i}\ L^i, \quad \phi_6 = \phi_{6,s}\ K^s, \quad \phi_8 = \phi_8^a\ t^a_{\mathbf 3}, \quad \phi_1 = \phi_1\ \mathbb{1}\,,
\end{align}
where
\begin{align}
	[L^i]^{jk} = L^{ijk} =\frac{1}{\sqrt 2} \epsilon^{i jk}, \quad t^a_{\mathbf 3} = \frac 12 \lambda^a
\end{align}
and 
\begin{align}
	&K_1 = \mqty( 1&0&0\\0&0&0\\0&0&0 ), \quad K_2 = \frac{1}{\sqrt 2} \mqty( 0&1&0\\ 1&0&0\\ 0&0&0 ), \quad K_3 = \mqty( 0&0&0\\ 0&1&0\\ 0&0&0 ), \\
	&K_4 = \frac{1}{\sqrt 2} \mqty(0&0&0\\ 0&0&1\\ 0&1&0), \quad K_5 = \mqty( 0&0&0\\ 0&0&0\\ 0&0&1 ), \quad K_6 = \frac{1}{\sqrt 2} \mqty( 0&0&1\\ 0&0&0\\ 1&0&0 ).
\end{align}
The matrices are normalised as follows: 
\begin{align}
	\Tr(L^i L_j) = \frac 12 \epsilon^{ikl} \epsilon_{jkl} =  \delta^i_{\phantom ij}, \quad \Tr(K^s K_t) = \delta^s_{\phantom st}, \quad \Tr(t^a_\mathbf{3} t^b_\mathbf {3}) = \frac 12 \delta^{ab}.
\end{align}
For the $\SU(6)/\SO(6)$ coset, the fields can be embedded within the symmetric and anti-symmetric two-index irrep as follows:
\begin{alignat}{2}
	\phi_{20} &= \frac{1}{\sqrt 2} \mqty( \phi_8 & \phi_6 \\ \phi_6^c & \phi_8^T ) &&\quad\Rightarrow \Tr(\phi_{20}^\dagger \phi_{20}) = \frac 12 \pi_8^a \pi_8^a +  \phi_{6,s}\, \phi_6^{c,s}, \\
	\phi_{15} &= \frac{1}{\sqrt 2} \mqty( \phi_8 + \frac{1}{\sqrt 6} \phi_1 & \phi_3^c \\ \phi_3 & -\phi_8^T - \frac{1}{\sqrt 6}\phi_1 ) &&\quad \Rightarrow \Tr(\phi_{15}^\dagger \phi_{15}) = \frac 12\phi_1 \phi_1 + \frac 12 \pi_8^a \pi_8^a +  \phi_{3,i}\, \phi_3^{c,i}. \label{eq:norm}
\end{alignat}

\subsection{CCWZ symbols}\label{app:ccwz}
In the first sector of the hidden-symmetry extended coset, the Maurer-Cartan form reads:
\begin{align}
	\Omega_{0,\mu} &= iU_0^\dagger (\partial_\mu - i\hat{g}_s \mathbf  G_\mu - i\hat{g}' \mathbf  B_\mu) U_0  \\
	&= -\frac{\sqrt 2}{f_0} \partial_\mu \boldsymbol \pi_0 + \frac{i}{f_0^2} [\boldsymbol \pi_0, \partial_\mu \boldsymbol \pi_0] + \frac{\sqrt 2}{3f_0^3} [\boldsymbol \pi_0, [\boldsymbol \pi_0, \partial_\mu \boldsymbol \pi_0]] + \cdots  \nonumber\\
	&\quad +\hat{g}_s \left( \mathbf  G_\mu - \frac{\sqrt 2i}{f_0} [\boldsymbol{\pi}_0, \mathbf  G_\mu] - \frac{1}{f_0^2} [\boldsymbol{\pi}_0, [\boldsymbol{\pi}_0, \mathbf  G_\mu]] +\cdots \right)\nonumber \\
	&\quad +\hat{g}' \left( \mathbf  B_\mu - \frac{\sqrt 2i}{f_0} [\boldsymbol{\pi}_0, \mathbf  B_\mu] - \frac{1}{f_0^2} [\boldsymbol{\pi}_0, [\boldsymbol{\pi}_0, \mathbf  B_\mu]] +\cdots \right),
\end{align}
where the dots indicate higher orders in the pNGB fields.
Reading off the components for $d_{0,\mu}$ and $e_{0,\mu}$, we find:
\begin{align}
	d_{0,\mu} &= -\frac{\sqrt 2}{f_0} \left( \partial_\mu \boldsymbol{\pi}_0 - i\hat{g}_s [\boldsymbol{\pi}_0, \mathbf  G_\mu] - i\hat{g}' [\boldsymbol{\pi}_0, \mathbf  B_\mu] \right)+ \frac{\sqrt 2}{3f_0^3} [\boldsymbol{\pi}_0, [\boldsymbol{\pi}_0, \partial_\mu \boldsymbol{\pi}_0]]  + \cdots \\
	&= -\frac{\sqrt 2}{f_0} D_\mu \boldsymbol{\pi}_0 + \frac{\sqrt 2}{3f_0^3} [\boldsymbol{\pi}_0, [\boldsymbol{\pi}_0, \partial_\mu \boldsymbol{\pi}_0]] + \cdots\\
	e_{0,\mu} &= \hat{g}_s \mathbf  G_\mu + \hat{g}' \mathbf  B_\mu + \frac{i}{f_0^2} [\boldsymbol{\pi}_0, \partial_\mu \boldsymbol{\pi}_0] - \frac{\hat{g}_s}{f_0^2} [\boldsymbol{\pi}_0, [\boldsymbol{\pi}_0, \mathbf  G_\mu]]  -\frac{\hat{g}'}{f_0^2} [\boldsymbol{\pi}_0, [\boldsymbol{\pi}_0, \mathbf  B_\mu]] + \cdots \\
	&= \hat{g}_s \mathbf G_\mu + \hat{g}'\mathbf B_\mu + \frac{i}{f_0^2} [\boldsymbol{\pi}_0, D_\mu \boldsymbol{\pi}_0] + \cdots 
\end{align}
For the second sector, containing the heavy spin-1 resonances, we find:
\begin{align}
	\Omega_{1,\mu} &= iU_1^\dagger (\partial_\mu - i\tilde g \boldsymbol {\mathcal V}_\mu - i\tilde g \boldsymbol {\mathcal A}_\mu) U_1  \\
	&= -\frac{\sqrt 2}{f_1} \partial_\mu \boldsymbol \pi_1 + \frac{i}{f_1^2} [\boldsymbol \pi_1, \partial_\mu \boldsymbol \pi_1] + \frac{\sqrt 2}{3f_1^3} [\boldsymbol \pi_1, [\boldsymbol \pi_1, \partial_\mu \boldsymbol \pi_1]] + \cdots  \nonumber\\
	&\quad + \tilde g \left( \boldsymbol {\mathcal V}_\mu - \frac{\sqrt 2i}{f_1} [\boldsymbol{\pi}_1, \boldsymbol {\mathcal V}_\mu] - \frac{1}{f_1^2} [\boldsymbol{\pi}_1, [\boldsymbol{\pi}_1, \boldsymbol {\mathcal V}_\mu]] +\cdots \right)\nonumber \\
	&\quad + \tilde g \left( \boldsymbol {\mathcal A}_\mu - \frac{\sqrt 2i}{f_1} [\boldsymbol{\pi}_1, \boldsymbol {\mathcal A}_\mu] - \frac{1}{f_1^2} [\boldsymbol{\pi}_1, [\boldsymbol{\pi}_1, \boldsymbol {\mathcal A}_\mu]] +\cdots \right). 
\end{align}
Reading off the two components:
\begin{align}
	d_{1,\mu} &= \tilde g \boldsymbol {\mathcal A}_\mu - \frac{\sqrt 2}{f_1} \partial_\mu \boldsymbol \pi_1 - \frac{\sqrt 2i \tilde g}{f_1} [\boldsymbol \pi_1, \boldsymbol {\mathcal V}_\mu] - \frac{\tilde g}{f_1^2} [\boldsymbol \pi_1, [\boldsymbol \pi_1, \boldsymbol {\mathcal A}_\mu]] + \frac{\sqrt 2}{3f_1^3} [\boldsymbol \pi_1, [\boldsymbol \pi_1, \partial_\mu \boldsymbol \pi_1]] +\cdots \\
	e_{1,\mu} &= \tilde g\boldsymbol {\mathcal V}_\mu  - \frac{\sqrt 2i \tilde g}{f_1} [\boldsymbol \pi_1, \boldsymbol {\mathcal A}_\mu] + \frac{i}{f_1^2} [\boldsymbol \pi_1, \partial_\mu\boldsymbol \pi_1] - \frac{\tilde g}{f_1^2} [\boldsymbol \pi_1, [\boldsymbol \pi_1, \boldsymbol {\mathcal V}_\mu]] + \cdots
\end{align}
The elements above are used as building blocks for the effective Lagrangian we used in the main text.

\subsection{Couplings to gluons}\label{app:gluons}

The couplings of the coloured resonances to QCD gluons stem from the kinetic terms of the elementary and composite states. For simplicity, we consider here only the couplings to vectors $\mathcal{V}_r$.
In the hidden symmetry approach, the gauge kinetic terms read
\begin{align}
	\mathcal L_\mathrm{gauge} = -\frac 12 \Tr \mathbf G_{\mu\nu} \mathbf G^{\mu\nu} -\frac 12 \Tr \boldsymbol {\mathcal F}_{\mu\nu} \boldsymbol {\mathcal F}^{\mu\nu}
\end{align}
before the mixing of the elementary gluon with $\mathcal V_8$. At zeroth order in all couplings, $\mathcal O(g^0)$, we have
\begin{align}\label{eq:appLg0}
	\mathcal L_{g^0} \supset -\frac 14 (\partial_\mu G_{\nu}^a - \partial_\nu G_{\mu}^a)^2 -\frac 14 (\partial_\mu \mathcal V_{8\nu}^a- \partial_\nu \mathcal V_{8\mu}^a)^2 - \frac 12 \big|\partial_\mu \mathcal V_{3\nu,i} - \partial_\nu \mathcal V_{3\mu,i}\big|^2. 
\end{align}
At linear order, $\mathcal{O} (g^1)$, there are derivative terms, which can be written as
\begin{align}
	\mathcal L_{g} = - \frac 12 \hat g_s \,& f^{abc}\, (\partial_\mu G^a_\nu - \partial_\nu G^a_\mu) G^b_\mu G^c_\nu \nonumber\\
	-4 i\tilde g \Big[ & (\partial_\mu \mathcal V_{8\nu}^a - \partial_\nu \mathcal  V_{8\mu}^a) \mathcal V_8^{b,\mu} \mathcal V_8^{c\nu} \Tr( t^a_{\mathbf 3} t^b_{\mathbf 3} t^c_{\mathbf 3} ) \nonumber\\
	&- (\partial_\mu \mathcal V_{8\nu}^a - \partial_\nu \mathcal V_{8\mu}^a) \mathcal V_{3,i}^{c\mu} \mathcal V_{3,j}^\nu \Tr(t_{\mathbf 3}^a L^i L^j)  \nonumber\\
	&- (\partial_\mu \mathcal V_{3\nu,i}^{c} - \partial_\nu \mathcal V_{3\mu,i}^{c})  	\mathcal V_{3,j}^\mu \mathcal V_8^{a,\nu} \Tr(L^i L^j t^a_{\mathbf 3}) + \hc \Big] 
\end{align}
with traces
\begin{align}
	\Tr(t^a_{\mathbf 3} t^b_{\mathbf 3} t^c_{\mathbf 3})  = \frac 14 (d^{abc} + i f^{abc}), \qquad
	\Tr(L^i L^j t^a_{\mathbf 3}) = \frac 12 [t^a_{\mathbf 3}]^{ij}.
\end{align}
The fully symmetric term with $d^{abc}$ falls out due to symmetry, hence leaving
\begin{align}
	\mathcal L_g&=- \frac 12 \hat g_s \, f^{abc}\, (\partial_\mu G^a_\nu - \partial_\nu G^a_\mu) G^b_\mu G^c_\nu + i \tilde g \Bigg[  \frac 12 (\partial_\mu \mathcal V_{8\nu}^a - \partial_\nu \mathcal V_{8\mu}^a) \mathcal V_8^{b,\mu} \mathcal V_8^{c,\nu} \, if^{abc} \nonumber\\
	& \quad - (\partial_\mu \mathcal V_{8\nu}^a - \partial_\nu \mathcal V_{8\mu}^a) \mathcal V_{3,i}^{c\mu} \mathcal V_{3,j}^\nu \, [t^a_{\mathbf 3}]_{ij} 
	 - \big((\partial_\mu \mathcal V_{3\nu,i}^{c} - \partial_\nu \mathcal V_{3\mu,i}^{c})  	\mathcal V_{3,j}^\mu \mathcal V_8^{a,\nu} [t^a_{\mathbf 3}]_{ij} + \hc \big) \Bigg].
	\end{align}
Finally, the $\mathcal O(g^2)$ terms read 
\begin{align}
	\mathcal L_{g^2}&= - \frac{\hat g_s^2}{4} \, f^{abc} f^{ade} \, G^b_\mu G^c_\nu G^{d,\mu} G^{e,\nu} -\tilde g^2 \Bigg[ 
		 \frac 14 \mathcal V_{8\mu}^a \mathcal V_{8\nu}^b \mathcal V_8^{c,\mu} \mathcal V_8^{d,\nu} \,  f^{abe} f^{cde} \nonumber \\
	&\quad+ i \, \mathcal V_{8\mu}^a \mathcal V_{8\nu}^b \mathcal V_{3,i}^{c\mu} \mathcal V_{3,j}^\nu \, f^{abc}\, [t^c_{\mathbf 3}]_{ij}  + (\mathcal V_{3\mu,j} \mathcal V_{8\nu}^a - \mathcal V_{3\nu,j} \mathcal V_{8\mu}^a) \mathcal V_{3,i}^{c\mu} \mathcal V_{8}^{b,\nu} \, [t^b_{\mathbf 3} t^a_{\mathbf 3}]_{ij} \Bigg].
\end{align}

We now take into account the octet mixing, redefining the fields to mass eigenstates (c.f. main text) as follows
\begin{align}
	G_\mu \stackrel{\mathrm{phys}}{\longrightarrow} \cos\beta_8\, G_\mu - \sin\beta_8\, \mathcal V_{8,\mu}, \qquad \mathcal V_{8,\mu} \stackrel{\mathrm{phys}}{\longrightarrow} \sin\beta_8 \, G_\mu + \cos\beta_8\, \mathcal V_{8,\mu},
\end{align}
and we only show the couplings involving two heavy vectors, which are phenomenologically relevant for pair production via QCD interactions. We hence neglect terms of $\mathcal O(\mathcal V^3)$ and $\mathcal O(\mathcal V^4)$.
The kinetic terms in \cref{eq:appLg0} remain unaffected by the field redefinition.
Instead, the $\mathcal O(g^1)$ terms lead to the following couplings (up to two $\mathcal{V}$ fields):
\begin{align}
	\mathcal L_g = &- \frac 12 g_s \, f^{abc}\, (\partial_\mu G^a_\nu - \partial_\nu G^a_\mu) G^{b\mu} G^{c\nu} \nonumber\\
	& - \frac 12 g_s \, f^{abc}\, (\partial_\mu G^a_\nu - \partial_\nu G^a_\mu) \mathcal V^{b\mu}_{8} \mathcal V^{c\nu}_{8} - g_s f^{abc} (\partial_\mu \mathcal V^a_{8\nu}- \partial_\nu \mathcal V^a_{8\mu}) G^{b\mu} \mathcal V^{c\nu}_8 \nonumber\\
	&- ig_s (\partial_\mu G_{\nu}^a - \partial_\nu G_{\mu}^a) \mathcal V_{3,i}^{c\mu} \mathcal V_{3,j}^\nu \, [t^a_{\mathbf 3}]_{ij} - g_s\big(i(\partial_\mu \mathcal V_{3\nu,i}^{c} - \partial_\nu \mathcal V_{3\mu,i}^{c})  	\mathcal V_{3,j}^\mu G^{a,\nu} [t^a_{\mathbf 3}]_{ij} + \hc \big).  
\end{align}
Note that there is no $\mathcal V_8$-$G$-$G$ coupling.
Finally we turn to the $\mathcal O(g^2)$ terms:
\begin{align}
	\mathcal L_{g^2} = &- \frac{g_s^2}{4} \, f^{abe} f^{cde} \, G^a_\mu G^b_\nu G^{c,\mu} G^{d,\nu} \nonumber\\
	& - \frac{g_s^2}{2} \, f^{abe} f^{cde} \, \big(G^a_{\mu} G^b_{\nu} \mathcal V^{c\mu}_8 \mathcal V_8^{d\nu} + G^a_{\mu} \mathcal V^b_{8\nu} G^{c\mu} \mathcal V_8^{d\nu}  + G^a_{\mu} \mathcal V^b_{8\nu} \mathcal V^{c\mu}_8 G^{d\nu}  \big) \nonumber\\
	& -i g_s^2 \, G_{\mu}^a G_{\nu}^b  \mathcal V_{3,i}^{c\mu} \mathcal V_{3,j}^\nu \, f^{abc}\, [t^c_{\mathbf 3}]_{ij}  - g_s^2(\mathcal V_{3\mu,j} G_{\nu}^a - \mathcal V_{3\nu,j} G_{\mu}^a) \mathcal V_{3,i}^{c\mu} G^{b,\nu} \, [t^b_{\mathbf 3} t^a_{\mathbf 3}]_{ij}.
\end{align}
We remark that couplings with three $\mathcal V$ and one gluon are also present, but they are only relevant for triple production:
\begin{equation}
    \mathcal L_{g^2} \supset -\frac 12 g_s (\tilde g c_8^3 - \hat g_s s_8^3) \, f^{abe} f^{cde} \, \big(G^a_{\mu} V^b_{8\nu} V^{c\mu}_8 V_8^{d\nu} + V^a_{8\mu} G^b_{\nu} V^{c\mu}_8 V_8^{d\nu} \big)
\end{equation}
and the coupling is not fully fixed by gauge invariance \cite{Zerwekh:2012bf}.

\subsection{Couplings to pNGBs}\label{app:decays}
We recall that the Lagrangian in unitary gauge reads:
\begin{align}
	\mathcal L =&-\frac{1}{2g_s^2} \Tr \mathbf  G_{\mu\nu} \mathbf  G^{\mu\nu} - \frac{1}{2g^{\prime\,2}} \Tr \mathbf  B_{\mu\nu} \mathbf  B^{\mu\nu} - \frac{1}{2\tilde g^2} \Tr \boldsymbol{\mathcal F}_{\mu\nu} \boldsymbol{\mathcal F}^{\mu\nu} \nonumber \\
	&+ \frac{f_0^2}{2} \Tr d_{0,\mu} d_0^{\mu} + \frac{f_1^2}{2} \Tr d_{1,\mu} d_1^{\mu} + rf_1^2\, \Tr d_{0,\mu} d_1^{\mu} \nonumber \\
	&+ \frac{f^2_K}{2} \Tr e_{0,\mu} e_0^{\mu} + \frac{f^2_K}{2} \Tr e_{1,\mu} e_1^{\mu} - f^2_K \Tr e_{0,\mu} e_1^{\mu} \nonumber \\
	&+ \mathcal L_\mathrm{fermions}\,,
\end{align}
where it is the $d^2$- and $e^2$-terms that contain couplings of the spin-1 resonances to the pNGBs.
It turns out that these interactions only come as two independent traces\footnote{$\Tr( \partial_\mu \boldsymbol \pi [\boldsymbol \pi, \boldsymbol { V}^\mu] ) = -\Tr( [\boldsymbol \pi, \partial_\mu \boldsymbol \pi] \boldsymbol { V}^\mu) $ and $\Tr( [\boldsymbol \pi, [\boldsymbol \pi, \partial \boldsymbol \pi]] \boldsymbol {\mathcal A} )  = - \Tr([\boldsymbol \pi, \partial\boldsymbol \pi] [\boldsymbol \pi, \boldsymbol {\mathcal A}]) = \Tr(\partial \boldsymbol \pi [\boldsymbol \pi, [\boldsymbol \pi, \boldsymbol {\mathcal A}]])$.}:
\begin{align}
	\mathcal O_V &= i \Tr([\boldsymbol \pi, \partial_\mu \boldsymbol \pi] \boldsymbol{ V}^\mu),\\
	\mathcal O_{\mathcal A} &= \Tr( [\boldsymbol \pi, [\boldsymbol \pi, \partial_\mu \boldsymbol \pi]] \boldsymbol {\mathcal A}^\mu ) ,
\end{align}
where $V=\mathcal V,G,B$ is a generic vector.
Both $\mathcal O_V$ and $\mathcal O_{\mathcal A}$ are hermitian. Traces with two pNGBs and one axial-vector vanish as they contain three broken generators of the coset.

When rotating the pNGBs to the physical eigenstates with \cref{eq:piredefinition}, $\pi_0$ and $\pi_1$ only differ in the prefactor:
\begin{equation}
	\pi_0 = \pi \, \frac{1}{\sqrt{1-R^2}}, \qquad \pi_1 = - \pi\, \frac{R}{\sqrt{1- R^2}}, \qquad \mathrm{with} \quad R=r \frac{f_1}{f_0},   
\end{equation}
where we switched to unitary gauge, $\pi_U\to0$ and $\pi_P\to\pi$.
In the operators $\mathcal O_V$ and $\mathcal O_{\mathcal A}$, it is therefore sufficient to keep track of the number of $\pi_0$ and $\pi_1$ fields:
\begin{align}
	\mathcal O_V^{k,l} &= \left( \frac{1}{\sqrt{1-R^2}} \right)^k \left( -\frac{R}{\sqrt{1-R^2}} \right)^l i \Tr([\boldsymbol \pi, \partial_\mu \boldsymbol \pi] \boldsymbol{ V}^\mu),\\
	\mathcal O_{\mathcal A}^{k,l} &= \left( \frac{1}{\sqrt{1-R^2}} \right)^k \left( -\frac{R}{\sqrt{1-R^2}} \right)^l \Tr( [\boldsymbol \pi, [\boldsymbol \pi, \partial_\mu \boldsymbol \pi]] \boldsymbol {\mathcal A}^\mu ) .
\end{align}

We can now collect the terms that facilitate the vector and axial vector decays, starting with the $d^2$-terms.
In the first sector we have
\begin{align}
	\frac{f_0^2}{2} \Tr(d_{0,\mu} d_0^{\mu}) &\supset \Tr(D_\mu\boldsymbol{\pi}_0 D^\mu\boldsymbol{\pi}_0) \\ 
	&\supset 2i g_s\Tr(\partial_\mu \boldsymbol{\pi}_0 [\boldsymbol{\pi}_0, \mathbf G^\mu]) + 2ig' \Tr(\partial_\mu \boldsymbol{\pi}_0 [\boldsymbol{\pi}_0, \mathbf B^\mu]) \\
	&= -2g_s\, \mathcal O_G^{2,0} - 2g'\, \mathcal O_B^{2,0},
\end{align}
Analogously, in the second sector we get
\begin{align}
	\frac{f_1^2}{2} \Tr(d_{1,\mu} d_1^{\mu}) &\supset 2i \tilde g \Tr(\partial_\mu \boldsymbol{\pi}_1 [\boldsymbol{\pi}_1 , \boldsymbol {\mathcal V}^\mu]) + \frac{\sqrt 2 \tilde g}{3f_1} \Tr([\boldsymbol \pi_1, [\boldsymbol \pi_1, \partial_\mu \boldsymbol \pi_1]] \boldsymbol {\mathcal A}^\mu) \nonumber\\
	&\quad + \frac{\sqrt 2\tilde g}{f_1} \Tr(\partial_\mu \boldsymbol \pi_1 [\boldsymbol \pi_1, [\boldsymbol \pi_1, \boldsymbol {\mathcal A}^\mu]]) \\ 
	&= 2i \tilde g \Tr(\partial_\mu \boldsymbol{\pi}_1 [\boldsymbol{\pi}_1 , \boldsymbol {\mathcal V}^\mu]) + \frac{4\sqrt 2 \tilde g}{3f_1} \Tr([\boldsymbol \pi_1, [\boldsymbol \pi_1, \partial_\mu \boldsymbol \pi_1]] \boldsymbol {\mathcal A}^\mu) \\
	&= -2\tilde g\, \mathcal O_{\mathcal V}^{0,2} + \frac{4\sqrt 2 \tilde g}{3f_1} \,\mathcal O_{\mathcal A}.
\end{align}
Next we have the mixed $d_0d_1$-term, which contributes
\begin{align}
	&rf_1^2\,\Tr d_{0,\mu} d_1^{\mu} \\
	&\supset r f_1^2 \, \Tr\!\Bigg( \frac{2i\tilde g}{f_0 f_1} \partial_\mu \boldsymbol \pi_0 [\boldsymbol \pi_1, \boldsymbol {\mathcal V}^\mu ] + \frac{\sqrt 2 \tilde g}{f_0 f_1^2} \partial_\mu \boldsymbol \pi_0 [\boldsymbol \pi_1, [\boldsymbol \pi_1, \boldsymbol {\mathcal A}^\mu]] \nonumber \\ 
	&\qquad \qquad \qquad - \frac{2i}{f_0 f_1} [ \boldsymbol \pi_0, (g_s \mathbf   G_\mu + g' \mathbf B_\mu)] \partial^\mu \boldsymbol \pi_1 + \frac{\sqrt 2 \tilde g}{3f_0^3} [\boldsymbol \pi_0, [\boldsymbol \pi_0, \partial_\mu \boldsymbol \pi_0]] \boldsymbol {\mathcal A}^\mu \Bigg) \\
	&= r f_1^2 \left( -\frac{2\tilde g}{f_0 f_1} \, \mathcal O_{\mathcal V}^{1,1}  + \frac{\sqrt 2 \tilde g}{f_0 f_1^2} \, \mathcal O_{\mathcal A}^{1,2} + \frac{2g_s}{f_0 f_1} \, \mathcal O_G^{1,1} + \frac{2g'}{f_0 f_1} \, \mathcal O_B^{1,1} + \frac{\sqrt 2 \tilde g}{3f_0^3} \, \mathcal O_{\mathcal A}^{3,0} \right).
\end{align}
Now on to the $e^2$-terms:
\begin{align}
	\frac{f_K^2}{2} \Tr e_{0,\mu} e_0^{\mu} &\supset \frac{f_K^2}{2f_0^2} \Tr( 2i (g_s \mathbf  G_\mu + g' \mathbf B_\mu) [\boldsymbol \pi_0 , \partial^\mu \boldsymbol \pi_0] ) = \frac{f_K^2}{f_0^2} \left( g_s\, \mathcal O_G^{2,0} + g' \, \mathcal O_B^{2,0} \right),
\end{align}
\begin{align}
	\frac{f_K^2}{2} \Tr e_{1,\mu} e^{\mu}_1 &\supset \frac{f_K^2}{2f_1^2} \Tr( 2i \tilde g \boldsymbol {\mathcal V}_\mu [\boldsymbol \pi_1, \partial^\mu \boldsymbol \pi_1] +\frac{2\sqrt 2 \tilde g}{f_1} [\boldsymbol \pi_1, \partial_\mu \boldsymbol \pi_1] [\boldsymbol \pi_1, \boldsymbol {\mathcal A}^\mu] ) \\
	&= \frac{f_K^2}{f_1^2} \left( \tilde g \, \mathcal O_{\mathcal V}^{0,2} - \frac{\sqrt 2 \tilde g}{f_1} \, \mathcal O_{\mathcal A}^{0,3}\right).
\end{align}
And finally $e_0 e_1$:
\begin{align}
	-f_K^2 \Tr e_{0,\mu} e_1^{\mu} &\supset -f_K^2 \Tr( \frac{i} {f_1^2} \mathbf V_\mu [\boldsymbol \pi_1 , \partial_\mu\boldsymbol \pi_1] + \frac{i\tilde g}{f_0^2} [\boldsymbol \pi_0, \partial_\mu \boldsymbol \pi_0] \boldsymbol {\mathcal V}^\mu + \frac{\sqrt 2\tilde g}{f_0^2 f_1} [\boldsymbol \pi_0, \partial_\mu\boldsymbol \pi_0] [\boldsymbol \pi_1 , \boldsymbol {\mathcal A}^\mu] ) \\
	&= -f_K^2 \left( \frac{g_s}{f_1^2} \, \mathcal O_G^{0,2} + \frac{g'}{f_1^2} \, \mathcal O_B^{0,2} + \frac{\tilde g}{f_0^2} \, \mathcal O_{\mathcal V}^{2,0} - \frac{\sqrt 2 \tilde g}{f_0^2 f_1} \, \mathcal O_{\mathcal A}^{2,1} \right).
\end{align}
Finally we take into account the $\mathcal V_8^\mu$-$G^\mu$ and the $\mathcal V_1^\mu$-$B^\mu$ mixings:
\begin{align}
	G_\mu \stackrel{\mathrm{phys}}{\longrightarrow} \cos\beta_8\, G_\mu - \sin\beta_8\, \mathcal V_{8,\mu}, \qquad \mathcal V_{8,\mu} \stackrel{\mathrm{phys}}{\longrightarrow} \sin\beta_8 \, G_\mu + \cos\beta_8\, \mathcal V_{8,\mu}
\end{align}
and analogous for the singlet. For the full vector multiplet, this means
\begin{align}
	\boldsymbol {\mathcal V}^\mu \stackrel{\mathrm{phys}}{\longrightarrow} \boldsymbol {\mathcal V}_{3/6}^\mu + \cos\beta_8\, \boldsymbol {\mathcal V}_8^\mu + \cos\beta_1\, \boldsymbol {\mathcal V}_1^\mu + \sin\beta_8\, \mathbf G^\mu + \sin\beta_1\, \mathbf B^\mu
\end{align}
where $\mathcal V_3$ ($\mathcal V_6$) contains both $\mathbf 3$ and $\mathbf {\bar 3}$ ($\mathbf 6$ and $\mathbf {\bar 6}$).
All in all, the decays into pNGBs are described by
\begin{align}
	\mathcal L_\mathrm{decays} = C_{\mathcal A} \mathcal O_{\mathcal A} + C_{\mathcal V_1}\mathcal O_{\mathcal V_1} + C_{\mathcal V_8}\mathcal O_{\mathcal V_8} + C_{\mathcal V_{3/6}}\mathcal O_{\mathcal V_{3/6}} 
 \label{eq:Ldecays}
\end{align}
with coefficients 
\begin{align}
	C_{\mathcal V_1} &= \frac{(r^2-1) f_{K}^{2} }{f_{0}^{2} \cdot \left(1 - R^{2}\right)} (\tilde g c_8 + g_s s_8) + 2\frac{1+R^2}{1-R^2} g_s s_8, \\
	C_{\mathcal V_8} &= \frac{(r^2-1) f_{K}^{2} }{f_{0}^{2} \cdot \left(1 - R^{2}\right)} (\tilde g c_8 + g_s s_8) + 2\frac{1+R^2}{1-R^2} g_s s_8, \\
	C_{\mathcal V_{3/6}} &=  \frac{\tilde g (r^2-1) f_K^2}{f_0^2 \cdot (1-R^2)} ,\label{eq:cv36app}\\
	C_{\mathcal A} &= \frac{\sqrt 2 \tilde g r}{3} (1-r^2) (f_1^2-3f_K^2).
\end{align}
We recall that the singlet will have additional mixing in the electroweak sector of the theory, which we do not include here.

Finally we have to calculate the operators $\mathcal O_{\mathcal V/\mathcal A}$. 
In the main text, we focus on the phenomenology of the $\mathcal V_8$, so we calculate $\mathcal O_{\mathcal V_8}$.
In the $\SU(6)/\SO(6)$ coset, 
\begin{align}
	\mathcal O_{\mathcal V_8} = \frac{i}{4\sqrt 2} \, \pi_8^a \lr {}^\mu \pi_8^b \, \mathcal V^c_{8,\mu} \, f^{abc} + \frac{1}{2\sqrt 2} \pi_{6,s} \lr {}^\mu \pi_6^{c,t} \, \mathcal V_{8,\mu}^a \, [t^a_{\mathbf 6}]_t^{\phantom ts},
\end{align}
while in the $\SU(6)/\Sp(6)$ coset we have a triplet pNGB,
\begin{align}
	\mathcal O_{\mathcal V_8} = \frac{i}{4\sqrt 2} \, \pi_8^a \lr {}^\mu \pi_8^b\, \mathcal V^c_{8,\mu} \, f^{abc} + \frac{1}{2\sqrt 2} \pi_{3,i} \lr {}^\mu \pi_3^{c,j} \, \mathcal V_{8,\mu}^a \, [t^a_{\mathbf 3}]_i^{\phantom ij}.
\end{align}
The operator
\begin{align}
	\mathcal L = \lambda\, \mathcal V_{8,\mu}^a \, \pi^x \lr{}^\mu \pi^y\, c^{axy}
\end{align}
with $\SU(3)$ tensor $c^{axy}$ yields 
\begin{align}
	\Gamma(\mathcal V_8 \to \pi_r \pi_r) = |\lambda_{8rr}|^2 \mathcal C_{\mathbf r} \, \frac{M_{\mathcal V_8}}{3\times 2^7\pi} \left(1 - \frac{4M_{\pi_r}^2}{M_{\mathcal V_8}^2}\right)^{3/2}
\end{align}
with colour factor $\mathcal C_{\mathbf r}=c^{axy} c^{ayx}$.
We have 
\begin{align}
	\lambda_{888} = \frac{iC_{\mathcal V_8}}{4\sqrt 2}, \qquad \lambda_{866} = \frac{C_{\mathcal V_8}}{2\sqrt 2} = \lambda_{833} 
\end{align}
and 
\begin{align}
	\mathcal C_{\mathbf r} = C_2(\mathbf r)  \dim(\mathbf r) \,\, \Rightarrow \,\, \mathcal C_{\mathbf 8} = 3\times8 = 24, \quad \mathcal C_{\mathbf 6} = \frac{10}{3}\times 6 =  20, \quad \mathcal C_{\mathbf 3} = \frac 43 \times 3 = 4,
\end{align}
and therefore 
\begin{align}
	\br(\mathcal V_8 \to \pi_8\pi_8) : \br(\mathcal V_8 \to \pi_6 \pi_6^c) : \br(\mathcal V_8 \to \pi_3 \pi_3^c) = 3:10:2,
\end{align}
assuming all scalars have the same mass.

\subsection{Couplings to top and bottom quarks via partial compositeness}\label{app:PC}

In the models under consideration \cite{Ferretti:2013kya}, the top mass is generated via partial compositeness, i.e.\ a linear mixing of the elementary top fields to composite baryons. In the hidden symmetry framework, baryons can be included as spin-1/2 resonances transforming under irreducible representations $R_\Psi$ of the hidden symmetry $\mathcal{G}'$, hence they couple to vector and axial resonances via their gauging. However, note that as $\mathcal{G}'$ is broken down to $\mathcal{H}'$, different components of $R_\Psi$ will have a different mass as generated by the strong dynamics.

In general, to provide a successful top mass generation, all models must contain baryons with the same quantum numbers as the left-handed and right-handed fields, $q_L$ and $t_R$. Hence all models contain at least two fields, $\Psi_Q$ and $\Psi_T$, both being vector-like. The mixing pattern with the elementary top fields \cite{Kaplan:1991dc} is such that the left-handed components of $\Psi_Q$ and the right-handed component of $\Psi_T$ have large mixing angles with the mass eigenstates, while the other two chiralities have mixing angles suppressed by the ratio of the electroweak scale over the Higgs decay constant, $v/f$. Hence, the couplings to the physical top and bottom fields can be obtained from the baryon couplings with the substitutions:
\begin{eqnarray}
    & \Psi_{T,L} \to s_R \frac{v}{f}\ t_L\,, \quad \Psi_{Q_u,L} \to s_L\ t_L\,, \quad  \Psi_{Q_d,L} \to s_L\ b_L\,, & \nonumber \\
    & \Psi_{T,R} \to s_R \ t_R\,, \quad \Psi_{Q_u,R} \to s_L \frac{v}{f}\ t_R\,; & 
\end{eqnarray}
where $s_{L/R} = \sin \theta_{L/R}$, with $\theta_{L/R}$ being two independent mixing angles. With this recipe, we can convert the couplings of vector and axial resonances to baryons into couplings to physical top and bottom. 

Independently on $R_\Psi$, the octets always couple to the baryons in all models, with vector-like couplings given by
\begin{equation}
    \mathcal{L}_\text{PC} \supset \tilde{g} \mathcal{V}_{8,\mu}^a \left[ \bar{\Psi}_Q \gamma^\mu t_3^a \Psi_Q + \bar{\Psi}_T \gamma^\mu t_3^a \Psi_T \right] + i \tilde{g} \mathcal{A}_{8,\mu}^a \left[ \bar{\Psi}_Q \gamma^\mu \gamma^5 t_3^a \Psi_Q + \bar{\Psi}_T \gamma^\mu\gamma^5 t_3^a \Psi_T \right]\,.
\end{equation}
As the mixing to the physical top fields are chiral, the effective couplings of the mass eigenstates can be parameterised as:
\begin{align}
    \mathcal{L}_\text{PC} \supset &\; \bar{t} \slashed {\mathcal V}_8^a t^a_{\mathbf 3} \left( g_{\rho t t, LL} P_L + g_{\rho t t, RR} P_R \right) t +  \bar{b} \slashed {\mathcal V}_8^a t^a_{\mathbf 3} \left( g_{\rho b b, LL} P_L + g_{\rho b b, RR} P_R \right) b + \nonumber \\
    &  \bar{t} \slashed {\mathcal A}_8^a t^a_{\mathbf 3} \left( - g_{a t t, LL} P_L + g_{a t t, RR} P_R \right) t +  \bar{b} \slashed {\mathcal A}_8^a t^a_{\mathbf 3} \left( - g_{a b b, LL} P_L + g_{a b b, RR} P_R \right) b \,,
\end{align}
where $P_{L/R}$ are the usual chiral projectors and, from the above substitutions, we have at leading order in $v/f$:
\begin{eqnarray}
    & g_{\rho tt,LL} = g_{\rho bb,LL} = \tilde{g} \cos \beta_8\ s_L^2\,, \quad g_{a tt,LL} = g_{a bb,LL} = \tilde{g} \ s_L^2 & \nonumber \\
    & g_{\rho tt, RR} = \tilde{g} \cos \beta_8\ s_R^2\,, \quad g_{a tt, RR} = \tilde{g} \ s_R^2\,, \quad g_{\rho bb, RR} = g_{a bb, RR} = 0\,, &
\end{eqnarray}
where the $\cos \beta_8$ factor comes from the mixing of $\mathcal{V}_8$ to gluons.

Regarding the non-octet resonances, $\mathcal{V}_3$ and $\mathcal{A}_6$, they can couple to a ditop state only when the underlying fermion $\chi$ carries baryon number $1/3$, i.e. in the case $\psi \psi \chi$. Hence, all baryons transform as the fundamental of $\SU(6)$. In this case, the baryons that mix with the elementary top and bottom can be embedded in the $6$ of $\SU(6)$ as:
\begin{equation}
    \Psi_6 = \begin{pmatrix} \Psi \\ \Psi^C  \end{pmatrix}\,.
\end{equation}
Hence, the triplet and sextet coupling to baryons will have the generic form:
\begin{equation}
    \mathcal{L}_\text{PC} \supset \tilde{g} \mathcal{V}_{3,\mu} \overline{\Psi^c} \gamma^\mu \Psi + i \tilde{g} \mathcal{A}_{6,\mu} \overline{\Psi} \gamma^\mu \Psi^c + \text{h.c.}
\end{equation}
with appropriate colour contractions. Taking into account the electroweak charges, only the singlet $\Psi_T$ is allowed such couplings, hence the effective couplings of $\mathcal{V}_3$ and $\mathcal{A}_6$ only involve tops and can be parameterised as
\begin{equation}
\mathcal{L}_\text{PC} \supset g_{\rho t t, LR}\ \overline{t^c} \slashed{\mathcal{V}}_3 t +  g_{a t t, LR}\ \overline{t} \slashed{\mathcal{A}}_6 t^c + \text{h.c.} 
\end{equation}
where
\begin{equation}
    g_{\rho tt, LR} = g_{a tt, LR} = \tilde{g} s_R^2 \frac{v}{f}\,.
\end{equation}
The $v/f$ suppression compared to the octet couplings stems from the fact that the couplings always involve one left-handed and one right-handed baryon, hence at least one (the left-handed) will have a suppressed mixing angle to the physical tops.

\bibliographystyle{utphys}
\bibliography{literature}

\end{document}